\documentclass[preprint,1p,times]{elsarticle}
\usepackage{setspace}
\usepackage{graphicx}
\usepackage{inputenc}
\usepackage{multirow}
\usepackage{tikz}
\usetikzlibrary{shapes.geometric, arrows}

\usepackage{caption}
\usepackage{subcaption}

\usepackage{placeins}

\usepackage{mathtools}
\newcommand\givenbase[1][]{\:#1\lvert\:}
\let\given\givenbase

\DeclarePairedDelimiterX\Basics[1](){\let\given\sgiven #1}

\usepackage{amssymb}
\usepackage{caption}
\usepackage{graphicx}
\usepackage{multirow}
\usepackage{algorithm}
\usepackage{mathtools}
\usepackage[noend]{algpseudocode}

\makeatletter
\def\algbackskip{\hskip-\ALG@thistlm}

\usepackage{algorithm}% http://ctan.org/pkg/algorithms
\usepackage{algpseudocode}% http://ctan.org/pkg/algorithmicx
\usepackage{amssymb}
\usepackage{bbold}
\usepackage{booktabs}
\usepackage{amsmath}
\usepackage{graphicx}
\usepackage{multirow}
\newcommand*{\Perm}[2]{{}^{#1}\!P_{#2}}%
\usepackage{url}
\usepackage{cite}
\usepackage{fancyhdr}
\newcommand{\ra}[1]{\renewcommand{\arraystretch}{#1}}

\usepackage{amsmath,amssymb,amsfonts}
\usepackage{textcomp}
\usepackage{xcolor}

\usepackage[normalem]{ulem}
\usepackage{ragged2e}
\usepackage{array}
\usepackage{graphicx}
\usepackage{booktabs}

\def\BibTeX{{\rm B\kern-.05em{\sc i\kern-.025em b}\kern-.08em
    T\kern-.1667em\lower.7ex\hbox{E}\kern-.125emX}}

\journal{Computer Speech \& Language}

\begin{document}

\begin{frontmatter}

\title{Cross-Corpora Spoken Language Identification with Domain Diversification and Generalization}

\author[mysecondaryaddress]{Spandan Dey\corref{corr1}}
\ead{sd21@iitkgp.ac.in}

\address[mysecondaryaddress]{Department of Electronics \& Electrical Communication Engineering, Indian Institute of Technology Kharagpur, Kharagpur, India-721302}

\author[mymainaddress]{Md Sahidullah\corref{corr2}}
\ead{md.sahidullah@inria.fr}

\address[mymainaddress]{Universit\'{e} de Lorraine, CNRS, Inria, LORIA, F-54000, Nancy, France}

\author[mysecondaryaddress]{Goutam Saha}
\ead{gsaha@ece.iitkgp.ac.in}

\cortext[corr1]{Corresponding author.}
\cortext[corr2]{Md Sahidullah is presently an Independent Researcher and Consultant.}
\begin{abstract}
This work addresses the cross-corpora generalization issue for the low-resourced spoken language identification (LID) problem. We have conducted the experiments in the context of Indian LID and identified strikingly poor cross-corpora generalization due to corpora-dependent non-lingual biases. Our contribution to this work is twofold. First, we propose \emph{domain diversification}, which diversifies the limited training data using different audio data augmentation methods. We then propose the concept of maximally diversity-aware cascaded augmentations and optimize the augmentation fold-factor for effective diversification of the training data. Second, we introduce the idea of \emph{domain generalization} considering the augmentation methods as pseudo-domains. Towards this, we investigate both domain-invariant and domain-aware approaches. Our LID system is based on the state-of-the-art \emph{emphasized channel attention, propagation, and aggregation based time delay neural network} (ECAPA-TDNN) architecture. We have conducted extensive experiments with three widely used corpora for Indian LID research. In addition, we conduct a final blind evaluation of our proposed methods on the Indian subset of VoxLingua107 corpus collected in the wild. Our experiments demonstrate that the proposed domain diversification is more promising over commonly used simple augmentation methods. The study also reveals that domain generalization is a more effective solution than domain diversification. We also notice that domain-aware learning performs better for same-corpora LID, whereas domain-invariant learning is more suitable for cross-corpora generalization. Compared to basic ECAPA-TDNN, its proposed domain-invariant extensions improve the cross-corpora EER up to $5.23\%$. In contrast, the proposed domain-aware extensions also improve performance for same-corpora test scenarios.

\end{abstract}
\end{frontmatter}
%\vspace{-.15cm}

% \tableofcontents

\section{Introduction}
\label{sec:intro}
%\vspace{-0.15cm}

In the era of voice assistants, smart homes, and numerous other innovative gadgets, the speech-based \emph{human-computer interaction} (HCI) is becoming a part of our daily life. Across the globe, the number of users relying upon modern-day speech applications, such as \emph{automatic speech recognition} (ASR), \emph{speaker recognition and verification},  \emph{speech synthesis}, \emph{speech emotion recognition} (SER), will keep on increasing~\citep{clark2019state}. For ease of use, the speech applications should operate efficiently on multiple input languages. A front-end \emph{language identification} (LID) block is generally used for multilingual applicability~\citep{radford2022robust}. The LID system predicts the language from the input, and the following speech application then adapts the mode of operation accordingly~\citep{li2013spoken}. 

For efficient real-world deployment of the speech applications, improving the generalization of the front-end LID module is important. For LID systems, the generalized classifier should be robust against several non-lingual sources, such as speaker identity, gender, age, dialects, and accents, mismatches due to channel and background environments~\citep{dey2021cross}. We can assume that diversity in non-lingual effects is expected to increase in larger speech corpora with greater diversity in data collection settings. Such corpora may improve the generalization by reducing the mismatch with unknown test utterances~\citep{sturm2014simple}. Therefore, the availability of a large and diversified corpus is important for developing robust LID systems.

Speech corpora collected through the initiatives, such as \emph{NIST LRE}~\citep{greenberg20122011,sadjadi20182017} and \emph{OLR challenges}~\citep{li2020ap20,tang2019ap19}, consist of large-scale multilingual speech corpora with carefully verified ground truths. These corpora are extensively used to develop and evaluate LID algorithms using deep neural network (DNN) architectures~\citep{gonzalez2015frame,lopez2016use,monteiro2019residual}, which substantially advanced LID research in the recent decade. However, many of the widely spoken languages in the world are still low-resourced due to the lack of standard large-scale corpora. The LID research with those low-resource languages is commonly conducted with independently created small-scale in-house corpora, which have limited non-lingual variations~\citep{dey2021overview}. As a consequence, training state-of-the-art DNN-based LID systems using such small corpus is prone to overfitting. To address this issue, researchers have experimented with different approaches, such as applying data augmentation, feature selection, regularizers, cross-validation, or model ensembling~\citep{dey2021cross,reddy2013identification,shen2017conditional}. Still, achieving generalization for the LID systems with low-resource languages remains the key challenge~\citep{dey2021cross}.

In this work, we aim to improve the generalization of the low-resourced LID systems under the \emph{cross-corpora} framework. In LID literature, most work considers the performance by training system on the train set and evaluating it on the dedicated evaluation set from the same corpus. However, for the low-resourced corpora, the data diversity within the dataset can be limited. As a consequence, even a decent performance on its own evaluation set may question its generalization on evaluations with different data. Hence, we explore cross-corpora evaluations to assess the generalization ability of the LID system. For each corpus, we independently train LID models and evaluate them with the test subsets for all the available corpora. Our preliminary investigation~\citep{dey2021cross} with different low-resourced LID corpora revealed that the cross-corpora LID performances are strikingly inferior compared to the same-corpora performances. In that study, we used vanilla x-vector time delay neural network (TDNN) architecture~\citep{snyder2018x} using \emph{mel frequency cepstral coefficients} (MFCC) features. The current work utilizes state-of-the-art \emph{emphasized channel attention, propagation, and aggregation in TDNN} (ECAPA-TDNN) architecture~\citep{desplanques2020ecapa} and advances this further for better generalization.

In this study, our key focus is to address the cross-corpora generalization by reducing the performance mismatch with the same-corpora evaluation. We propose novel domain diversification strategies and further frame the cross-corpora LID problem by introducing the \emph{domain generalization} (DG) concepts~\citep{cha2021domain} in this field. One possible way to address cross-corpora performance issues is by following the \emph{domain adaptation} (DA)~\citep{pan2010domain,zhang2020unsupervised} approaches. In the DA framework, the training corpus is used as the source domain, and we have access to limited labeled or unlabeled data from the evaluation condition, which we refer to as the target domain. In the cross-corpora scenario, similarly, the training-set of one corpus can be considered as the source domain, and the training-set utterances from the remaining corpora can be considered as individual target domains. However, applying DA methods in a straightforward way is not possible in some practical scenarios for two reasons. Firstly, we may not have any specific target domain audio data at all while training the LID systems. Secondly, even if we have some target domain data (which can be the other corpora except the training corpus in our problem), applying DA can improve the generalization on that specific target corpus, and the generalization to unseen conditions will remain a key concern. Therefore, we apply the concepts of DG in this work. This is a more recent and practical solution, and it does not require specific target domain data~\citep{zhou2022domain}. Rather, in the DG framework, multiple pseudo-target domains are created from the source domain, and the model is trained to achieve better generalization across multiple pseudo-domains. We create different pesudo-domains following diverse domain diversification methods. These pesudo-domains are then utilized to develop multiple DG approaches. We propose domain-invariant and domain-aware DG approaches for learning language-discriminating knowledge by extending the state-of-the-art ECAPA-TDNN architecture.

\begin{figure*}[!t]
% \vspace{-.6cm}
    \centering
    \includegraphics[trim=0.22cm 2cm 0.25cm 
    1cm,clip,width=\textwidth]{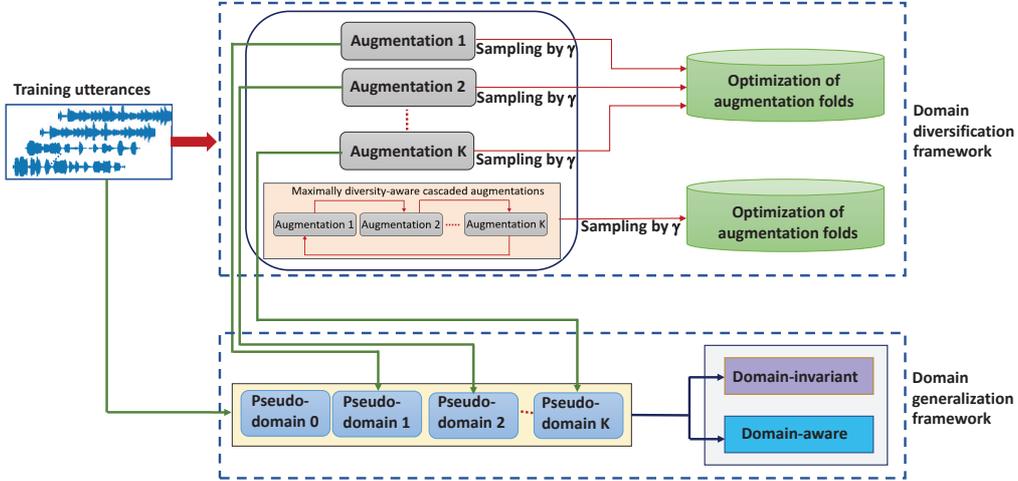}
    \vspace{-.3cm}
    \caption{Overview of the proposed frameworks for improved cross-corpora generalization in LID. Here, $\gamma$ denotes the augmentation fold-factor (explained in Section~\ref{sec:6_c} and $K$ denotes the total number of signal-level augmentation categories.}
    \label{fig:overall}
    \vspace{-.3cm}
\end{figure*}

In Fig.~\ref{fig:overall}, we have presented the overview of our proposed solutions towards better generalization of LID systems. Fig.~\ref{fig:overall} presents the main two approaches that we have explored: domain diversification and domain generalization. 

The major contributions of this work are summarized as follows:
%\vspace{-0.1cm}
\begin{itemize}

    \item We explore several audio data augmentations, including commonly used methods as well as several new ways of increasing the data diversity. The comprehensive study of different augmentation techniques could also be useful for other speech-based classification tasks.  
    
    \item To maintain a trade-off between the amount of original data and augmented data, we conduct a systematic study to analyze the impact of augmentation \emph{fold-factor} for the generalization of TDNN-based LID classifiers. The analysis also provides an optimal augmentation fold-factor in terms of computation complexity and generalization performance. 

    \item We further increase the diversity of the low-resourced corpora by proposing the \emph{maximally diversity-aware cascaded augmentation}. It generates more realistic augmented audio data based on maximum mutual domain diversity measures while preserving the perception and intelligibility of the spoken language.
    
  \item Considering the different domain-diversified audio categories as multiple pseudo-domains, we introduce the concept of DG for improving the cross-corpora performance. We extend the state-of-the-art ECAPA-TDNN architecture by introducing two opposite but functionally similar DG approaches suitable for the unknown target domains. The first one is domain-invariant approach, where the LID model is trained to get confused in the task of identifying different pseudo-domain. In the domain-invariant approach, we explore adversarial training and \emph{maximum mean discrepancy} (MMD) approaches. In the second domain-aware approach, along with learning the language discriminating cues, we train the LID model to recognize the different pseudo-domains. We explore multitask learning with pseudo-domain classification as an additional task in the domain-aware approach.
  
  \item We further extend the conventional MMD-based DG approach by proposing the \emph{multi-domain} MMD (MD-MMD) algorithm. Instead of considering the pseudo-domains altogether as a single target domain, the proposed method computes diversity-weighted multi-domain MMD losses and jointly optimizes the combined MD-MMD losses.
  
  \item We conduct systematic benchmarking and analysis of the same-corpora and cross-corpora LID performances with three independently collected LID corpora in the Indian languages. Finally, we perform a blind evaluation of the algorithms in a fourth unseen corpus.
\end{itemize}
%\vspace{-0.1cm}

The rest of the paper is organized as follows: Section~\ref{Sec:2} formally introduces the cross-corpora evaluation paradigm. Section~\ref{Sec:4} describes the corpora and the experimental methodology used for this work. The proposed domain diversification techniques for improving the cross-corpora performance are discussed in Section~\ref{Sec:5}. The cross-corpora problem is then addressed from domain generalization approaches in Section~\ref{Sec:6}. In Section~\ref{Sec:7}, we present the experimental results. We provide additional analysis of the cross-corpora problem, including blind evaluations on an unseen in-the-wild corpus in Section~\ref{Sec:8}. Finally, we conclude the paper in Section~\ref{Sec:9} by providing a summary and recommendations.

\section{Problem definition \& related works}
\label{Sec:2}
% \vspace{-0.15cm}

In this section, we have formally introduced our cross-corpora evaluation protocol and discussed its applications in various speech-based classification tasks.

\subsection{Formulation of cross-corpora evaluation}
%\vspace{-0.1cm}
In the cross-corpora evaluation paradigm, the model is trained on one corpus and tested on multiple corpora. If the test samples are taken from the hold-out part of the training corpus, then it is called \emph{same-corpora} evaluation. When the test samples are taken from other corpora, it is called \emph{cross-corpora} evaluation. Next, we present a formal definition of the cross-corpora evaluation followed by the perspectives and constraints.

\textbf{Definition}: Let us consider a cross-corpora experimental setting with $C$ number of corpora. Let the data of the $i$-th corpus be denoted as $\{(\mathbf{X}_k^{i},y_k^{i})\}_{k=1}^{N^i}$, where $i\in[1,C]$ and $N^i$ denoting the number of samples in corpus $i$. $\mathbf{X}_k$ is the features of the $k$-th utterance of the $i$-th corpus, and $y_k$ is the corresponding label. 

For each corpus $i$, the feature space $\mathcal{X}^i = [\mathbf{X}^i_1, \mathbf{X}^i_2, \cdots, \mathbf{X}^i_{N_i}]$ is partitioned into disjoint training and evaluation parts $\mathcal{X}^i_T$ and $\mathcal{X}^i_E$. The corresponding label space $\mathcal{Y}^i = [y^i_1, y^i_2, \cdots, y^i_{N_i}]  \in \{1, 2, \cdots, L\}$ ($L$ denotes the total number of target classes) is also partitioned as $\mathcal{Y}^i_T$ and $\mathcal{Y}^i_E$ respectively. 

Let us consider a mapping function $f(\mathbf{X};\mathbf{W}): \mathbf{X} \longrightarrow \mathbf{y}$ with $\mathbf{W}$ denoting the learnable parameter of the mapping function. Associated with $f$, the corresponding multi-class classification loss function is denoted as $\mathcal{L}(\mathbf{y},f(\mathbf{X};\mathbf{W}))$, which needs to be minimized during training.

Consider, a model is trained with the training set from the $i$-th corpus, $(\mathcal{X}_T^{i},\mathcal{Y}_T^{i}
)$ and is tested with the evaluation set from the $j$-th corpus, $\mathcal{X}^{{j}}_E$.  \textcolor{black}{For the test utterances}, the posterior prediction probability \textcolor{black}{vector} for this train-test pair is denoted as \textcolor{black}{$p \left ( \mathcal{Y}^{j}_E \given \mathcal{X}^{j}_E;\mathbf{W}^{i} \right )$}, where \textcolor{black}{$\mathbf{W}^{i}$} denotes the learned parameters trained on \textcolor{black}{$\mathcal{X}^{i}_T$}. For \textcolor{black}{$j=i$}, the evaluation follows the \emph{same-corpora} configuration. Our key interest is in the \emph{cross-corpora} conditions, i.e., when \textcolor{black}{$j \neq i$}.
%\textcolor{black}{ We then use a performance metric evaluator $g(\mathbf{y}^{j}_E,p(\mathbf{y}^{j}_E|\mathbf{X}^{j}_E;\mathbf{W}^{i}))$ that takes the true class labels ($\mathbf{y}$) and predicted labels ($p(\mathbf{y}^{j}_E|\mathbf{X}^{j}_E;\mathbf{W}^{i})$) as input and computes a scalar value indicating the classification performance.} 
\textcolor{black}{We then consider a performance metric computation function $g$, that takes the true class labels ($\mathcal{Y}$) and predicted labels ($p(\mathcal{Y}\given \mathcal{X};\mathbf{W})$) as inputs and computes a scalar value indicating the classification performance.}

% \textcolor{black}{For the standalone LID model trained with the training set of corpus $i$, the objective of the cross-corpora evaluation is to reduce the performance mismatch with the same-corpora evaluation metric when . We express the performance mismatch as:}

\textcolor{black}{Consider, $(\mathcal{X}_T^{i},\mathcal{Y}_T^{i})$ is used to train a standalone LID model ($f^i$) trained with corpus $i$. The key objective of cross-corpora evaluation is to assess and minimize the performance mismatch by $f^i$ when it is evaluated on $\mathcal{X}^{i}_E$ and $\mathcal{X}^{j}_E$, respectively. The cross-corpora performance mismatch (for corpora $i$ and $j$) is formulated as:}

\begin{equation}
\textcolor{black}{
D_g(i,j) = \left| g\left(\mathcal{Y}^{i}_E,p(\mathcal{Y}^{i}_E \given \mathcal{X}^{i}_E;\mathbf{W}^{i})\right) - g\left(\mathcal{Y}^{j}_E,p(\mathcal{Y}^{j}_E \given \mathcal{X}^{j}_E;\mathbf{W}^{i})\right)\right|
}
\label{eq:cc_basic}
\end{equation}

\textcolor{black}{The aim of this work is to minimize the performance mismatch ($D_g(i,j), \forall \; \{i,j\} \in [1,C]$) without degrading the same-corpora performance.} We can achieve this by estimating the parameters of the LID model ($\mathbf{W}^{i}$) such that the following condition is finally satisfied. 

\begin{equation}
\textcolor{black}{
\min\limits_{\mathbf{W}^{i}\;\given\;\mathcal{X}_T^{i}} \mathcal{L}\left(\mathcal{Y}^{i}_E,f(\mathcal{X}^{i}_E;\mathbf{W}^{i})\right)\; \mbox{s.t} \; D_g(i,j) \to 0 
\label{eq:cond}
}
\end{equation}

\textcolor{black}{Eq.~\ref{eq:cond} is needed to be ensured $\forall \; \{i,j\} \ni j \neq i$.}

\textbf{Perspectives \& constraints}: 
While deploying the trained standalone models in the real-world, test samples can be from any unseen domains. Therefore, generalizing over specific target domains may not be the most efficient solution for such cases. Our goal is to improve generalization over unseen domains with the evaluation performance on multiple corpora as a reference.  Hence, apart from the training corpus, utterances from the remaining corpora is not utilized as target domains.
In our evaluation framework, we apply the following constraints considering the challenges in the real-world application of the LID systems.

\begin{itemize}
    \item Independent models \textcolor{black}{$(f^{i})$} are trained using only the training data from a single corpus \textcolor{black}{$(\mathcal{X}^{i}_T,\mathcal{Y}^{i}_T)$, where $i\in[1,C]$}. Utterances from the training samples of other corpora $\mathcal{X}^{j}_T \ni j \neq i$ are not used during the system development.

      \item \textcolor{black}{For improving cross-corpora performance mismatch between train corpus $i$ and evaluation corpus $j$, we prohibit using the feedback from the same-corpora evaluation of the LID model trained with corpus $j$. This is to ensure the generalization are not biased towards corpus $j$.}
    
    \item For a fair performance comparison, we follow a closed-set LID problem with a fixed set of languages. All the corpora share the common label space: \textcolor{black}{$\mathcal{Y} = \{ \bigcap\mathcal{Y}^{i}\}_{i=1}^{C}$}.
    
\end{itemize}

\textcolor{black}{One may consider the cross-corpora study as an application of the commonly practiced domain adaptation (DA) methods used in the machine learning literature~\citep{pan2010domain,zhang2020unsupervised}. Following Eq.~\ref{eq:cond}, under the DA formulation of the cross-corpora study, corpora $i$ and $j$ can be considered as source and target domains, respectively. However, there is a fundamental difference between the objective of our cross-corpora generalization and the goal of the DA approaches. DA is effective for scenarios when the target data is known prior. However, we are interested in improving LID generalization during real-world deployments where the test data is unknown and can come from diverse non-lingual domains. For such \emph{out-of-domain} generalization, the formulation of our cross-corpora evaluation task is more appropriate. Unlike DA, we do not use subset of utterances from corpus $j$ as target domain data to compute $\mathbf{W}^{i}$. Rather, we follow domain generalization (DG)~\citep{zhou2022domain} (which is fundamentally similar to DA), where instead of using specific target domains, we create multiple pseudo-domains from the source corpus $i$ using several domain diversification methods. The pseudo-domain data are then used to compute $\mathbf{W}^{i}$.}

\subsection{Related works on cross-corpora evaluations}

The majority of the research in cross-corpora evaluation is so far conducted for computer vision applications, including object detection~\citep{khosla2012undoing,deng2018image}, facial expression recognition~\citep{zhu2015transfer}, deepfake detector~\citep{nadimpalli2022improving}, person re-identification~\citep{wang2018transferable}. In speech processing applications, cross-corpora evaluations are often adopted to assess the robustness, especially for tasks, such as emotion recognition and speaker recognition anti-spoofing, where dataset variability is limited due to the constraints in data collection.

In the speech emotion recognition (SER) field, the study reported in~\citep{schuller2010cross} performed cross-corpora SER with speaker-wise feature normalization using six different corpora. Cross-corpora SER performance is severely degraded except when the test data has background similarities with the train data. In~\citep{zhang2011unsupervised}, the authors used six emotional speech corpora and improved cross-corpora performance by adding unlabeled emotional utterances to the agglomerated training data. Several other works in SER also investigated cross-corpora evaluations~\citep{vlasenko2014modeling,vlasenko2013parameter,toledo2013voice,singh2021non,EmoNet}.

In speaker recognition anti-spoofing, the work in~\citep{paul2017} conducted cross-corpora experiments with two corpora, ASVspoof2015 and BTAS2016, which covered common spoofing attacks, speech synthesis, and voice conversion. This work demonstrated poor generalization in different data conditions. In~\citep{korshunov2019cross}, the authors used two anti-spoofing corpora, ASVspoof and AVspoof, and improved cross-corpora generalization by combining multiple systems with score fusion. Another recent study with seven seven anti-spoofing corpora demonstrated different levels of generalization with various classifiers~\citep{chettri2021data}. The study further correlated cross-corpora performance with different signal-level descriptors. 

Cross-corpora evaluations were also explored to detect the presence of dysarthria in speech~\citep{gillespie2017cross}, music mood prediction~\citep{7395312}, multi-modal voice activity detection (VAD)~\citep{9133504}, speech enhancement~\citep{pandeyself2022}, and speech recognition~\citep{tsakalidis2005acoustic}. In all cases, the authors demonstrated poor generalization due to large mismatch in data conditions.

\textcolor{black}{Studies with cross-corpora evaluations are limited in language recognition research compared to other speech-related tasks discussed above. A few studies included cross-corpora experimental results even though it was not explicitly mentioned in the papers~\citep{alumae22_odyssey,ferrer2022discriminative,liu22e_interspeech,liu2022efficient,liu2022enhancing,valk2021voxlingua107}. The work in~\citep{valk2021voxlingua107} trained the LID systems on subsets of VoxLingua107 and evaluated them on  KALAKA-3 and LRE07. They evaluated the robustness of different statistical and neural network models for cross-corpora LID. However, they did not conduct additional modifications to improve the cross-corpora generalization. Another recent work~\citep{ferrer2022discriminative} proposed a discriminative hierarchical model for language recognition and used a combination of NIST LRE, RATS, NIST SRE, VoxLingua107, Panarabic, CALLHOME and CALLFRIEND databases for training and a combination of LRE15, LRE17, LASRS, KALAKA and Crowdsource (CROWD) databases for evaluation. The aim of this work was to develop a discriminative probabilistic linear discriminant analysis (PLDA) framework as a better alternative to the commonly used generative modeling. The study in~\citep{liu22e_interspeech} incorporated hierarchical phonotactical information using multitask learning in LID. They evaluated their proposed algorithms on the NIST LRE17 and AP17-OLR databases independently. The LRE17 evaluation utterances are sampled from narrow-band MLS14 and wide-band VAST databases, which were not used in training and development in LRE17. In~\citep{liu2022efficient}, the authors separately used the NIST LRE17 and AP17-OLR corpora and removed the redundancies in the pre-trained multilingual wav2vec2.0 XLSR model based speech representations by using squeeze-excitation (SE) and linear bottleneck modules. In~\citep{alumae22_odyssey}, the authors used a custom training set for the unconstrained LID task of the OLR21 challenge. They combined training utterances of OLR21 with the Mozilla Common Voice~\citep{mozilla} data. The authors then used the pre-trained wav2vec 2.0 model and fine-tuned it using the custom training set.}

\textcolor{black}{Most of the above works violated the cross-corpora evaluation standards by mixing data from multiple corpora, and their key objectives were not focused on explicitly reducing the cross-corpora performance mismatch. Moreover, they mainly focus on major languages with rich linguistic resources. Our current work focuses on improving cross-corpora language recognition using databases that are developed independently and contain limited resources.}

\section{Experimental setup}
\label{Sec:4}
This section describes our experimental setup, including the details of speech corpora, feature extraction, and neural network architecture.

\subsection{Description of corpora}
We consider three corpora that are most widely used in the Indian LID literature. They are: \emph{IIITH-ILSC} (IIITH)~\citep{vuddagiri2018iiith}, \emph{LDC 2017S14} (LDC)~\citep{LDC}, and \emph{IITKGP-MLILSC} (KGP)~\citep{maity2012iitkgp}. For our experiments, we have selected the five languages, \emph{Bengali, Hindi, Punjabi, Tamil, and Urdu}, which are common to all three corpora. In Table~\ref{tab:corpora_comparison}, we have summarized the meta-data of the three corpora.

%\vspace{-.1cm}
\begin{table}[!htbp]
\centering
\caption{Meta-data comparison of the three speech corpora used in our experiments. We have shown the collective information for all the common subsets of the five languages considered for this study. (BN: Broadcast news, CTS: Conversational telephone speech.)}
\begin{scriptsize}
%\vspace{-.1cm}
%\vspace{-0.1cm}
\label{tab:corpora_comparison}
\centering
\renewcommand{\arraystretch}{1.1}
\resizebox{.8\textwidth}{!}{%
\centering
\begin{tabular}{|@{}| l| l| l| l| @{}| }
\hline
\textbf{Meta-data}                     & \textbf{IIITH}            & \textbf{LDC} & \textbf{KGP} \\ 
\hline
\textcolor{black}{\textbf{Duration}}       & \textcolor{black}{15 hours}                    & \textcolor{black}{118 hours}              & \textcolor{black}{4 hours}               \\ 
\textcolor{black}{\textbf{Speakers}} & \textcolor{black}{250}                             & \textcolor{black}{378}                      & \textcolor{black}{67}                     \\ 
\textcolor{black}{\textbf{Utterances}}              & \textcolor{black}{12,663}                             & \textcolor{black}{577}                        & \textcolor{black}{155}                     \\ 
\textbf{Mode of speech}    & BN and CTS              & CTS                      & BN              \\ 
\textbf{Accent \& dialect}   & Mostly standard & Variations exist & Standard \\
\textbf{Environment}             & Studio, real-world & Real-world                & Studio        \\ 
\textbf{Gender ratio} & Balanced  & Not maintained  & Balanced \\
\textbf{Audio channel} & Mono & Stereo & Mono \\
\textbf{Audio format}                  & 16~kHz (\texttt{.wav})                   & 8~kHz (\texttt{.flac})             & 8~kHz (\texttt{.wav})            \\ 
\hline
\end{tabular}}
% \vspace{-0.1cm}
\end{scriptsize}
\end{table}

The meta-data comparison in~Table~\ref{tab:corpora_comparison} lists the differences and similarities among the three corpora. The duration of speech indicates that these are low-resource corpora compared to the LID corpora in major languages. Out of these three, the KGP is the smallest one, with just four hours of speech data, whereas the LDC is the largest one, with about $24$ hours of audio data per language. IIITH and KGP data are mostly collected from broadcast news (BN). The LDC data contains conversational telephone speech (CTS) \textcolor{black}{recordings from the NIST LRE 2011 dataset}. The accents and dialects are also varied across the corpora. For example, the Bengali utterances present in the IIITH and KGP corpora contain dialects spoken mostly in India, whereas the LDC corpus contains dialects spoken in different parts of Bangladesh.

The IIITH and KGP corpora have dedicated evaluation protocols including data-split for training and test set. For LDC, there was no prior partitioning of the data. we randomly sample $80\%$ of the audio recordings as training data and the remaining are used for test. LDC utterances are stereo audio with each channel corresponding to one of the two speakers in the telephone call. For each language, one speaker is common in all the telephone call recordings. To ensure speaker disjoint partitions, we only consider the audio channel excluding the common speaker in the LDC test set. For all corpora, the training utterances are further partitioned with an $80:20$ ratio into speaker-disjoint training and validation sets.

\subsection{Data pre-processing \& feature extraction}
\textcolor{black}{Firstly}, we process the corpora to have a consistent sampling rate of 8~kHz, and an audio format of 16-bit PCM (\texttt{.wav}). Then we remove the non-speech regions using an energy-based voice activity detector (VAD)~\citep{benyassine1997silence}. We finally compute 20-dimensional \emph{mel frequency cepstral coefficients} (MFCCs) with 20 filters, and this is followed by \textcolor{black}{utterance-level} cepstral mean subtraction (CMS). For MFCC extraction, we use Hamming window of 25~ms and hop-length of 10~ms. \textcolor{black}{While the \emph{mel filter banks} (MFB) feature is frequently used with the neural network based classifiers, several works~\citep{kang2022deep,kumawat2021applying,snyder2018spoken} have also applied MFCCs as input to the TDNN-based classifiers as used in our work. Note that the existing work with the three used corpora~\citep{vuddagiri2018iiith,mandava2019attention,mandava2019investigation,chakraborty2021denserecognition} also reported their performances mostly using MFCCs.} We segment the training speech utterances with an active speech of 3~s duration to create the training examples. \textcolor{black}{For utterances with a duration of more than 3~s, we segment them into multiple chunks and discard the remaining portion. If the utterance duration is less than 3~s (which is very rare), we discard the utterance.} The test utterances are also similarly segmented with a duration of 3~s.

%\vspace{-0.1cm}
\subsection{Architecture description}
%\vspace{-0.1cm}
%\subsubsection{Baseline TDNN}

\emph{\textbf{Baseline TDNN}}: For the baseline reference, we use the \emph{x-vector TDNN}~\citep{snyder2018x} architecture trained on MFCC (with CMS post-processing) feature following our preliminary investigation in~\citep{dey2021cross}. The model consists of five TDNN layers followed by a \textcolor{black}{statistics pooling} (Stat pooling) layer and two fully connected (FC) layers~\citep{snyder2018spoken}. 
Table~\ref{tab:baseline_tdnn} shows the architectural details. 

\begin{table}[!htbp]
%\vspace{-.3cm}
%\ra{1.1}
\centering
\caption{Architecture details of the baseline x-vector TDNN with input feature dimension $F$, temporal dimension $T$, and the number of classes $L$.}
\label{tab:baseline_tdnn}
\resizebox{.8\linewidth}{!}{%
\begin{tabular}{|l|c|c|c|} 
\hline
 Layer & Context & Input dimension & Output dimension \\ 
\hline
 TDNN-ReLU& $\textcolor{black}{[t-2:t+2]}$ & $F*5 \times T$ & $512 \times T$ \\
 TDNN-ReLU& $\{t-2,t,t+2\}$ & $512*3 \times T$ & $512 \times T$ \\
 TDNN-ReLU& $\{t-3,t,t+3\}$ & $512*3 \times T$ & $512 \times T$ \\
 TDNN-ReLU& $\{t\}$ & $512 \times T$ & $512 \times T$ \\
 TDNN-ReLU& $\{t\}$ & $512 \times T$ & $1500 \times T$ \\
 Stat pooling & $T$ & $1500*T \times 1 $ & $3000 \times 1$ \\
 FC-ReLU& - & $3000 \times 1$ & $512 \times 1$ \\
 FC-ReLU& - & $512 \times 1$ &  $512 \times 1$\\
 FC-Softmax& - & $512 \times 1$ & $L \times 1$ \\
\hline
\end{tabular}
}
\end{table}

\emph{\textbf{ECAPA-TDNN}}: We have utilized the ECAPA-TDNN architecture~\citep{desplanques2020ecapa}, which outperformed other basic TDNN architectures for speaker verification tasks in~\citep{desplanques2020ecapa}. This is also very recently adopted for language recognition tasks~\citep{kang2022deep,thienpondt2022tackling}. The ECAPA-TDNN model modifies the basic x-vector architecture in different ways. Instead of the TDNN layers, three squeeze-excitation (SE) based residual blocks (SE-Res2) are used. Multi-layer feature aggregation (MFA) is performed with the outputs of these three blocks. The aggregated output is then fed to a channel and context attentive pooling layer (Attn Pooling) followed by a fully connected layer. We use batch normalization (BN) for 1d- convolutional (Conv1D) layers and used a drop-out of 0.25 for the linear layers. Table~\ref{tab:ECAPA_tdnn} presents a detailed description of this architecture.

\begin{table}
%\vspace{-.3cm}
%\ra{1.1}
\centering
\caption{Architecture details of the ECAPA-TDNN with \textcolor{black}{input feature dimension $F$, temporal dimension $T$, and number of classes $L$. (Here, BN denotes batch normalization).}}
\label{tab:ECAPA_tdnn}
\resizebox{.82\linewidth}{!}{%
\begin{tabular}{|l|c|c|c|} 
\hline
 Layer & Skip connect & Input dimension & Output dimension \\ 
\hline
 Conv1D+ReLU+BN & - & $F\times$T& $512\times$T\\
   SE-Res2-Block&-&512$\times$T&$512\times$T\\
  SE-Res2-Block&-&512$\times$T& $512\times$T \\
  SE-Res2-Block&-&512$\times$T&$512\times$T \\
   Conv1D+ReLU&layer 2,3,4 &$512*3\times$T&1536$\times$T\\
  Attn Pooling+BN&-&$1536*T\times$1&3072$\times$1 \\
  FC-ReLU&-&$3072\times$1&$192\times$1 \\
   FC-Softmax&-&$192\times$1& L$\times$1\\
\hline
\end{tabular}
}
%\vspace{-.4cm}
\end{table}

The TDNN architectures are implemented using the PyTorch library~\citep{paszke2019pytorch} of Python. We use a batch size of 32, AdamW optimizer~\citep{loshchilov2018decoupled}, \emph{reduce on plateau} based learning rate scheduler with an initial value of 0.001. We set the maximum epoch as 30 and apply early stopping criteria with five epochs on the validation loss. For the loss function, we have used the \emph{angular margin} (AM) softmax~\citep{wang2018additive} \textcolor{black}{with scale parameter as $30$ and margin as $0.2$.} 

\textcolor{black}{The neural LID models can be employed in two different ways. The first is an end-to-end architecture that builds one single module with speech as input and language labels as output~\citep{gonzalez2015frame}. The other type trains an embedding extractor first, which is followed by a backend classifier based on PLDA or cosine similarity~\citep{snyder2018spoken}. The end-to-end system is relatively simple and straightforward without the burden of training additional back-end for scoring. On the other hand, the additional backend-based approach is more flexible, especially when tackling out-of-set languages~\citep{martinez2011language}. Our present work focuses on closed-set LID problems with fixed five languages, and for brevity, we consider end-to-end architecture.}

\subsection{Performance evaluation metrics}
% \vspace{-0.1cm}
Following the literature, we use \emph{equal error rate} (EER) \citep{brummer2006application} and \emph{cost average} ($C_{\mathrm{avg}}$) as performance evaluation metrics. EER is defined as the error rate at the decision threshold where the false acceptance and false rejection rates are equal. \textcolor{black}{For multi-class scenarios, we compute one-vs-all EERs for each target language and report the average value.} The $C_{\mathrm{avg}}$ is used as the primary evaluation metric in the NIST language recognition evaluations (LRE)~\citep{sadjadi20182017} and OLR challenges~\citep{li2020ap20}. It is defined as follows \citep{sadjadi20182017}:
%\vspace{-0.05cm}
% \begin{equation} C_{avg}=\frac{1}{N}\sum\limits_{L_{t}}\begin{Bmatrix} P_{Target}\cdot P_{Miss}(L_{t})\\ +\sum\limits_{L_{n}}P_{Non-target}\cdot P_{FA}(L_{t}, L_{n}) \end{Bmatrix} 
% \end{equation}

\begin{equation} \textcolor{black}{C_{\mathrm{avg}}=\frac{1}{N}\sum\limits_{L_{t}}\left[ p_{\mathrm{Target}}\cdot p_{\mathrm{Miss}}(L_{t}) +\sum\limits_{L_{n}}p_{\mathrm{Non}-\mathrm{target}}\cdot p_{\mathrm{FA}}(L_{t}, L_{n})\right]}
\end{equation}

where $L_t$ and $L_n$ are the targets and non-target languages. $p_{\mathrm{Miss}}$ and $p_{\mathrm{FA}}$ are the probability of miss and false alarm, respectively. $p_{\mathrm{Target}}$ is the prior probability of the target languages, and it is set at $0.5$. $p_{\mathrm{Non}-\mathrm{target}}=(1-p_{\mathrm{Target}})/(L-1)$, where, $L$ is the total number of languages. The lower value of EER and $C_{\mathrm{avg}}$ indicates better language recognition performance. 

\section{Cross-corpora LID: Domain diversification}
\label{Sec:5}
% \vspace{-0.15cm}

Low-resourced LID corpora contain a substantially lesser number of utterances. Mostly the data collection source and configurations are very limited. So, the amount of non-lingual diversities in terms of varying backgrounds, recording instruments, room environments, and speaker populations are limited. These create corpora-specific non-lingual biases across the databases leading to inferior cross-corpora performances~\citep{dey2021cross}. Therefore, this section explores extensive audio augmentation methods to diversify the low-resourced corpora. In the speech processing literature, audio augmentation techniques, such as noise and non-speech addition, volume, pitch, and speed perturbations, are commonly used to artificially increase the amount and variations in the training data~\citep{mushtaq2020environmental,salamon2017deep,wei2020comparison}. For addressing the cross-corpora generalization problem, we focus on the extensive application of augmentation-based approaches for two reasons: (i)~increasing the training data helps avoiding overfitting issues~\citep{snyder2018spoken}, and (ii)~diversifying the training data can help reducing the corpora mismatch. To incorporate diverse sources of randomness in the training data, we explore several existing and new audio augmentation techniques. We further propose a new concept called \emph{maximally diversity-aware cascaded augmentation} approach and optimize the \emph{augmentation fold-factor}.

%The detailed description of the applied augmentation techniques are presented next.} 

%\vspace{-.1cm}
\subsection{Description of the explored augmentation methods}
\label{Sec:5.1}

Based on the manner of perturbing the audio data, we have classified the explored augmentation techniques into eight categories. We primarily focus on signal-level augmentation categories that operate directly on the raw waveform. Further, we have also explored feature-level augmentations. For each category, there are different sub-categories. The different augmentation categories are chosen based on the corpora-analysis conducted in our preliminary investigation~\citep{dey2021cross}, and they introduce diversities that help in reducing the corpora mismatch. We first studied the impact of each category on the LID performances. Then, we have pooled and sampled utterances from the different categories to accumulate the different ways of domain diversification they possess. We summarize the applied augmentation categories in Table~\ref{tab:aug_info}. A detailed description of these augmentation categories is presented as follows. 

\subsubsection{Additive non-speech augmentation}
The \emph{long-term average spectrum} (LTAS) and signal-to-noise ratio (SNR) histogram analysis in our preliminary investigation indicated that the channel effects and backgrounds may vary across different low-resourced LID corpora~\citep{dey2021cross}. Following this,
non-speech effects, such as additive noise, babble noise, and music clips, are added \textcolor{black}{to} the training utterances. This augmentation category is followed from~\citep{snyder2018x} and implemented with the Kaldi tool's~\citep{povey2011kaldi} VoxCeleb recipe~\footnote{\url{https://github.com/kaldi-asr/kaldi/tree/master/egs/voxceleb/v2}}. This uses non-speech samples from the MUSAN corpus~\citep{snyder2015musan}. %\textcolor{black}{Kaldi’s recipe also uses the RIR dataset for synthesizing reverberation effect. We have considered it in a different augmentation technique.}

\subsubsection{Signal parameter perturbations}
The utterances across the corpora can also vary based on the speaking style. For example, pitch, speaking rate, and loudness can be different across the speakers. 
To reduce the corpora mismatch in such perspectives, we randomly alter different signal parameters using the audio degradation toolbox~\citep{matthias2013a}.
\begin{itemize}
    \item Pitch: Random shifted by $[-4,4]$ semitones without changing the utterance lengths.
    
    \item Shift: Training utterances are randomly segmented into two parts and swapped.

    \item Speed: The speed of the playback is changed by randomly choosing $\Gamma \in [-15,+15]$. The length of the augmented audio ($l$) becomes $l'=((100-\Gamma)/100)l$. $\Gamma >0$ increases playback speed and $\Gamma <0$ reduces it~\citep{matthias2013a}.
    
    \item Volume: Random gains in the range of $[-30,40]$ dB.%, volume of the utterances is changed.
\end{itemize}
\begin{table*}[!t]
\centering
\caption{Summary of the different signal-level and feature-level augmentation categories and their respective sub-categories applied in the LID framework.}
%\vspace{-0.05cm}
\label{tab:aug_info}
\resizebox{\textwidth}{!}{%
\begin{tabular}{|l|l|l|l|l|l|l|l|} 
\hline
\begin{tabular}[c]{@{}l@{}}Additive non-speech\\augmentation ($A^{1}$)\end{tabular}      & \begin{tabular}[c]{@{}l@{}}Signal parameter\\perturbations ($A^{2}$)\end{tabular} & \begin{tabular}[c]{@{}l@{}}Bandwidth\\augmentation ($A^{3}$)\end{tabular}     & \begin{tabular}[c]{@{}l@{}}Environmental\\variations ($A^{4}$)\end{tabular}                                                   & \begin{tabular}[c]{@{}l@{}}Speech enhancement \\~ ~ ~ ~ ~  ($A^{5}$)\end{tabular}          & \begin{tabular}[c]{@{}l@{}}Audio encodings \\~ ~ ~ ($A^{6}$)\end{tabular}     & \begin{tabular}[c]{@{}l@{}}Restoration from \\lossy codecs ($A^{7}$)\end{tabular} & \begin{tabular}[c]{@{}l@{}}Feature-level\\augmentation ($A^{8}$)\end{tabular}  \\ 
\hline
\begin{tabular}[c]{@{}l@{}}Babble\\Music\\Noise\end{tabular} & \begin{tabular}[c]{@{}l@{}}Pitch\\Shift\\Speed\\Volume\end{tabular}          & \begin{tabular}[c]{@{}l@{}}Lower cutoff\\Upper cutoff\\Telephone-\\-channel BPF\end{tabular} & \begin{tabular}[c]{@{}l@{}}Smart phone \\Live hall \\Radio broadcast\\Vinyl recording\end{tabular} & \begin{tabular}[c]{@{}l@{}}Spectral subtraction\\MMSE-based\\De-reverberation\end{tabular} & \begin{tabular}[c]{@{}l@{}}a-law\\ima-adpcm\\oki-adpcm\\u-law\end{tabular} & \begin{tabular}[c]{@{}l@{}}\texttt{.aac}\\\texttt{.gsm}\\\texttt{.mp3}\\\texttt{.ogg}\\\texttt{.opus}\\\texttt{.wma}\end{tabular} & \begin{tabular}[c]{@{}l@{}}SpecAug\\Mixup\end{tabular}                     \\
\hline
\end{tabular}
}
%\vspace{-.42cm}
\end{table*}
\subsubsection{Bandwidth augmentation}
Following the LTAS comparison in~\citep{dey2021cross}, we observe that the spectrums of different corpora may vary prominently in the different frequency ranges. Hence, we process the utterances with filters of random cutoff frequencies.

\begin{itemize}
    \item Upper cutoff augmentation: Filtered with a fixed lower cutoff frequency $(f_L)$ of $20$~Hz. The upper cutoff frequencies $(f_U)$ are randomly selected $f_U \in [2500,3500]$~Hz.
    
    \item Lower cutoff augmentation: Filtered with a fixed  $f_U=f_s/2$  ($f_s$ denotes sampling frequency) and random  $f_L \in [50,200]$~Hz.
    
    \item Telephone channel band-pass filtering (BPF): To replicate the filtering effects that occurred in the old analog telephone channels (following the LDC meta-data of Table~\ref{tab:corpora_comparison}), we have filtered the audios using a band-pass filter with fixed $f_L = 300$~Hz and variable $f_U \in [3000,4000]$~Hz.
\end{itemize}

\subsubsection{Convolving different room environment responses}
Following Table~\ref{tab:corpora_comparison}, the IIITH and KGP utterances are mostly collected in recording studios. The LDC data is recorded outdoors and in smaller rooms. For simulating the different recording environments, utterances are convolved with randomly chosen room impulse responses from the RIR corpus following the Kaldi tool~\citep{povey2011kaldi}. Further, using the audio degradation toolbox~\citep{matthias2013a}, we transform the utterance environment into vinyl recordings, live hall recordings, smartphone playback, and radio broadcasting. 
%The impulse response of vinyl player is estimated using inverse filtering from an iZotope Vinyl 1.73b plugin. Playbacks from the Google Nexus One smartphone model is used to estimate the corresponding impulse responses. For the radio broadcasting, the utterances are transformed with dynamic range control (DRC).

\subsubsection{Speech enhancement}
The SNR histogram analysis in~\citep{dey2021cross} shows that the utterances across the corpora can be corrupted with different levels of noise. To reduce the mismatch due to the varying noise levels, we propose speech enhancements as an augmentation category. The Voicebox tool~\citep{brookes1997voicebox} is used to randomly apply one of the three speech enhancement algorithms to the utterances: spectral subtraction~\citep{berouti1979enhancement}, $\log$ MMSE noise estimation based enhancement~\citep{gerkmann2011unbiased}, and de-reverberation~\citep{doire2016single}.  

\subsubsection{Altering audio encoders}
%The telephone channel recordings of the LDC and the boradcasts of KGP were collected more than a decade ago. 
The intermediate data recording, processing, and storage across the corpora may not follow the same encoding format. Following the augmentations in~\citep{garcia2020magneto}, we explore several encoding formats for data augmentation. We have randomly converted the encoding format of the utterances among \emph{linear pulse code modulation} (L-PCM), a-law, u-law, Ima, and Oki adaptive delta PCM (ADPCM).\footnote{\url{https://github.com/FFmpeg/FFmpeg}}

\subsubsection{Restoration from lossy audio codecs}
% Table~\ref{tab:corpora_comparison} shows that the audio codecs followed in the three corpora are \texttt{.wav} and \texttt{.flac}. Both of them are lossless codecs. 
Lossy codecs allow degrading the audio fidelity up to some range for data compression. During intermediate processings, the utterances across the corpora could be stored in different lossy codec formats and then converted into lossless \texttt{.wav} (for IIITH and KGP) or \texttt{.flac} (for LDC). Augmentation by lossy codecs diversifies utterances by incorporating codec-dependent partial information losses. Using the SoX library,~\footnote{\url{https://github.com/rabitt/pysox}} we randomly convert the utterances in different lossy codecs, such as \texttt{.aac}, \texttt{.mp3}, \texttt{.gsm}, \texttt{.opus} and then restored them to \texttt{.wav}. 

\subsubsection{Feature-level augmentation}
The augmentation methods discussed so far are applied at the signal level. We also explore the suitability of two widely used augmentation methods which are applied at the feature level.

\emph{SpecAug}~\citep{park2019specaugment}: This feature-level augmentation method randomly modifies the filter-bank features by (i)~warping along the time axis, (ii)~time masking, and (iii)~frequency masking along random time-frequency locations. It increases the robustness against deformation along the time axis and partial information loss in specific frequency channels and time frames~\citep{park2019specaugment}. \textcolor{black}{SpecAug is primarily designed for spectrograms or MFBs. Due to masking, the standard SpecAug zeros out some of the MFB values, and the remaining values remain the same. However, in this work, we utilize the SpecAug on MFCCs. We do this by applying SpecAug directly on the MFBs that are generated as an intermediate stage during MFCC extraction. During the decorrelation by the final DCT stage, SpecAug perturbs the entire feature matrix. In this work, we utilize SpecAug with MFCCs mainly due to a fair comparison with the baseline and the existing LID literature with the three corpora which used MFCCs. Nevertheless, several earlier works also applied SpecAug on MFCCs~\citep{du2021data,rossenbach2020generating,tong2021asv,xia2021self}}.

\emph{Mixup}~\citep{zhang2017mixup}: This augmentation method generates new training example $(\mathbf{X}_{\mathrm{mix}},y_{\mathrm{mix}})$ by linear interpolation of randomly selected training examples $(\mathbf{X}_i,y_i)$ and $(\mathbf{X}_j,y_j)$. This is defined as:
\begin{equation}
    \begin{array}{l} {\mathbf{X}_{\mathrm{mix}}} = \theta {\mathbf{X}_i} + \left( {1 - \theta } \right){\mathbf{X}_j}, \\ {y_{\mathrm{mix}}} = \theta {y_i} + \left( {1 - \theta } \right){y_j}, \end{array} 
\end{equation}
where $\mathbf{X}_i$ and $\mathbf{X}_j$ denote the feature inputs of two independent examples, $y_i$ and $y_j$ are the corresponding labels presented in one-hot encoding. $\theta$ is the mixing parameter, and it follows beta-distribution where two shape parameters are equal and lie between 0 and $\infty$. Mixup creates soft labels $y_{\mathrm{mix}}$, which increase the robustness against corrupt labels and adversarial examples~\citep{zhang2017mixup}.

%\vspace{-.1cm}
\subsection{Optimizing augmentation fold}
\label{Sec:6c}
The data augmentation methods help in improving the generalization of neural network architectures such as TDNN~\citep{snyder2018x}. Studies have demonstrated that data augmentation with several methods helps in training larger models~\citep{garcia2020magneto}. \textcolor{black}{However, we suspect that to solve the issue of insufficient training data for the low-resourced corpora, if we use excessively large numbers of augmented samples, it can increase the computation. Further, with too many augmented utterances compared to the original samples, the LID model can start learning the acoustic variations and diverge from learning the true language-discriminating cues present in the original training data~\citep{sturm2014simple}.} In spite of notable advancements for exploring different audio augmentation techniques, we observe that the existing literature~\citep{garcia2020magneto,iqbal2021enhancing,wei2020comparison} does not provide comprehensive analysis on the amount of audio data to be augmented for training. In this work, we extensively explore finding the optimum (with respect to the same-corpora and cross-corpora LID performance) fraction of the augmented utterances to the original utterances in the combined (augmented + original) training set.

\textcolor{black}{Suppose, for any $i$-th corpus, the total number of non-augmented original training utterances be $N^i_O$. For each signal-level augmentation category $A^{k}$, where $k\in\{1, 2, \cdots, K\}$, we have created one augmented instance for all of its $S^k$ sub-categories. Hence, the total number of augmented utterances corresponding to $A^k$ is $M^k=S^k N^i_O$. If we randomly sample ${\eta}^k$ fractions from these $M^k$ utterances, then the total augmented utterances ($N^i_A$) are $N^i_A = \sum_{k=1}^{K}M^k{\eta}^k$.} 

If ${\eta}^k=\gamma/(KS^k)$ (where $\gamma$ is a non-zero positive real number), the total utterances in the combined augmented training set ($N^i_{\mathrm{Tot}}$) is:
\begin{equation}
\textcolor{black}{N^i_{\mathrm{Tot}}= N^i_O + N^i_A = (1 +\gamma) N^i_O}    
\end{equation}

 In this work, $\gamma$ is termed as the \emph{augmentation fold-factor}. \textcolor{black}{It denotes the ratio of augmented utterances to the original non-augmented training utterances in the combined training set.} We create different combined training sets for $\gamma = 0.5, 1, 2, 3, \mbox{and}~ 4$. Apart from evaluating the LID performance for each augmentation category, we also train independent LID models for different augmentation fold-factors and analyze their performances.

%\vspace{-.1cm}
\subsection{Cascading augmentation categories}
The domain diversification approach we follow aims to replicate the non-lingual diversities present in the unseen cross-corpora evaluation utterances to improve generalization. For this, we create more realistic augmented audios that simulate the different real-world acoustic scenarios.
Therefore, beyond the conventional ways of independently exploring the different audio augmentation categories, we propose an algorithm called  \emph{maximally diversity-aware cascaded augmentations} that applies multiple data augmentation categories together with the categories being selected based on the introduced diversity.

In the literature, several augmentation approaches include multiple signal processing stages to perturb the data. For example, the room environment-based augmentations ($A^{5}$) includes reverberation effect followed by noise additions. However, such examples are limited, and their potentialities are very little explored. In this work, we present one of the very first theoretical and experimental studies to propose efficient ways of cascading multiple augmentation categories. In the cascaded augmentation, the signal processing steps of categories $A^{i}$ and $A^{j}$ can be cascaded to form a new augmentation category $A^{\{i,j\}}$. Considering the diverse signal processing operations of the explored augmentation categories, the order of cascading is assumed to be non-commutative, $A^{\{i,j\}} \neq A^{\{j,i\}}$. We can pipeline two or more different augmentation categories to generate diverse augmented audio data. Here, we have to deal with a trade-off situation, the amount of diversification of the utterances versus preserving the semantic language labels of the augmented utterances. We address this trade-off by limiting the cascading to only two signal-level augmentation categories. 

Our method relies on the distances between the utterance embeddings for different augmentation categories. We train the LID model and extract the embeddings $Z^{k}$ of the training utterances belonging to the signal-level augmentation categories $A^{k} \; \forall\;k\in[1,K]$. \textcolor{black}{We use the output from the linear layer before the final softmax layer to extract the 192-dimensional utterance-level embeddings ($Z^{k}$).} Then, we calculate the \emph{diversity} among the embedding distributions $\mathbf{D}^{\{i,j\}}=\mathbf{D}||p(Z^{i}),p(Z^{j})||$. Here, $i,j \in [1,K]$ denote two different signal-level augmentation categories and $i \neq j$. 

To measure the diversity, we use three metrics~\citep{zhuang2020comprehensive}: \emph{cosine distance}, \emph{KL divergence}, and \emph{symmetric KL divergence}. For all the three distance metrics, we compute the embedding distances $(\mathbf{D}^{\{0,k\}})$ between the original utterances (denoted as $A^{0}$) and augmented utterances from $A^{k}$. We then perform correlation analysis between the $\mathbf{D}^{\{0,k\}}$ values and respective $c^{k}_{\mathrm{avg}}$ scores. Here,  $c^{k}_{\mathrm{avg}}$ denotes the same-corpora LID performance (on validation data) when the training data is formed by combining augmented utterances of category $A^{k}$ with the original training utterances. Based on experimental results (see Section~\ref{sec:7_e}), we select the KL divergence as diversity measure metric because it shows the most negative correlation values. The negative correlation implies that the performance becomes better (i.e., lower $C_{\mathrm{avg}}$) with higher domain diversity. Further, unlike the other two distance metrics, KL divergence is not symmetric, $\mathbf{D}_{\mathrm{KL}}^{\{i,j\}} \neq \mathbf{D}_{\mathrm{KL}}^{\{j,i\}}$. This is also concurrent with the non-commutative property of the cascaded augmentation approach. 

\textcolor{black}{For $K$ augmentation categories, there could be a total of $(\Perm{K}{2})$ cascaded pairs. We avoid repeating the same augmentation category. Therefore, the effective number of ways is $(\Perm{K}{2})-K$.} Correspondingly, we have one pair-wise KL divergence score $\mathbf{D}_{KL}^{\{i,j\}}$, where, $i,j \in [1,K]$ and $i \neq j$. The KL divergence scores between the embedding of two different augmentation categories are an indication of the variations in diversity they contribute to the training data. To maximize the diversity gain for the cascaded augmentations, for each augmentation category $i \in [1,K]$, we cascade the augmented category $j \in [1,K] \neq i$ such that: $\mathbf{D}_{\mathrm{KL}}^{\{i,j\}} > \mathbf{D}_{\mathrm{KL}}^{\{i,m\}}$, $\forall\; m \in [1,K] \neq j$. The generated cascaded augmented utterances are denoted as $A^{\{i,j\}}$. The summary of our proposed approach for generating the maximally diversity-aware cascaded augmented utterances is presented in Algorithm~\ref{tab:synthetic_aug_algo}.

%\begin{tiny}
\begin{algorithm}[!t]
\footnotesize
\ra{1}
\caption{Maximally diversity-aware cascaded augmentation.}\label{tab:synthetic_aug_algo}
\begin{algorithmic}[1]
\Procedure{}{Cascade $A^{i}$ and $A^{j}$ to $A^{\{i,j\}}$}\Comment{$i,j \in [1,K]$ and $i\neq j$}
\For {$i=1:K$}
\State $Z^{i} \gets A^{i}$\Comment{Z : TDNN embedding}
\For {$m=1:K$}
\If {$i \neq m$}
\State $\mathbf{D}_{\mathrm{KL}}^{\{i,m\}} \gets \mathbf{D}_{\mathrm{KL}}||p(Z^{i}),p(Z^{m})||$ \Comment{$\mathbf{D_{\mathrm{KL}}}$ : KL divergence}
\EndIf
\EndFor
\If {$\mathbf{D}_{\mathrm{KL}}^{\{i,j\}} > \mathbf{D}_{\mathrm{KL}}^{\{i,m\}}$, $\forall \:m \in [1,K] \neq j$}
%\Comment{$j \in [1,K]$}
\State Cascade $A^{i}$ followed by $A^{j}$.
\EndIf
\EndFor
\EndProcedure
\end{algorithmic}
\end{algorithm}

%\end{tiny}

% \vspace{-.2cm}
We also cascade two feature-level augmentation techniques, SpecAug, and mixup, independently on top of different signal-level augmentation methods. We perform this because the sources of audio perturbations covered by the signal and feature-level augmentations are different. Hence, cascading them can further enhance the diversity in the training data.

\section{Cross-corpora LID: Domain generalization}
\label{Sec:6}
Cross-corpora mismatches in non-lingual factors can be considered as a \emph{domain shift} problem~\citep{cha2021domain}. Domain shift caused by co-variance shifts, concept shifts, changes the feature-label joint distribution between the training (source) and testing (target) data samples, \textcolor{black}{$p(\mathcal{X}^{\mathrm{src}}\mathcal{Y}^{\mathrm{src}}) \neq p(\mathcal{X}^{\mathrm{tar}}\mathcal{Y}^{\mathrm{tar}})$}~\citep{moreno2012unifying}. In this section, we address the cross-corpora domain shift problem from domain generalization~\citep{gulrajani2020search} perspectives. 

%\vspace{-0.1cm}
\subsection{Domain adaptation vs. domain generalization}
\label{sec:6_a}
%\vspace{-0.1cm}
A common approach followed in the machine learning literature is to incorporate information from the target domain samples during training to mitigate the domain shift problem. The model, trained on the source domain, is usually adapted to that particular target domain~\citep{sarfjoo20_interspeech}. This approach is known as the \emph{domain adaptation}~\citep{zhang2020unsupervised,wang2018deep}. In our context, we can consider the training corpus as the source domain and the remaining two corpora as individual target domains. However, following the constraints of the cross-corpora evaluations presented in Section~\ref{Sec:2}, we refrain from using any data of the target domains. Moreover, adapting to any specific target domain does not guarantee generalization for any unseen test domains during real-world deployments~\citep{ding2017deep}. Considering these factors, we have selected an alternative approach of leveraging information from the source domain and improving the generalization of the LID models. This approach is called \emph{domain generalization}~\citep{zhou2022domain}. The concept of DG was first formally introduced in the computer vision field~\citep{blanchard2011generalizing}. It is gradually becoming an attraction in the research community for a \textcolor{black}{realistic solution to the domain shift problem~\citep{cha2021domain} when, rather than any specific target domain, we want to achieve robustness in real-world unseen conditions.} 

%\vspace{-0.1cm}
\subsection{Pseudo-domains: Solving for multiple source domains}
\label{sec:6_b}
%\vspace{-0.1cm}
In speech processing literature, the DG approaches are less explored due to the difficulties in collecting multi-source labeled data~\citep{gideon2021improving}. Hence, we first propose a solution for availing data from multiple source domains into our framework. In Section~\ref{Sec:5.1}, we have discussed the different signal-level augmentation categories,  applied for domain diversification. Here, we propose to utilize those augmentation categories as well as the original condition as multiple source domains for solving cross-corpora LID from a DG perspective. We have termed those synthetic domains as \emph{pseudo-domains} since they are artificially created from the source domain data. This approach of utilizing audio augmentations in DG is more appropriate for our current study with limited training data from low-resourced languages.  

Let us denote each pseudo-domain as \textcolor{black}{$\mathcal{D}^{\mathrm{dom}}=\{(\mathbf{X}_i^{\mathrm{dom}},y_i^{\mathrm{dom}})\}_{i=1}^{N_{\mathrm{dom}}}$}, where \textcolor{black}{$\mathrm{dom} \in [0,K]$} with $K$ denoting the total number of signal-level augmentation categories, \textcolor{black}{$N_{\mathrm{dom}}$} denoting the number of training samples for the corresponding pseudo-domain. The utterances \textcolor{black}{$\{(\mathbf{X}_i^{\mathrm{dom}},y_i^{\mathrm{dom}})\}$} are assumed to be sampled from the domain-specific joint distribution \textcolor{black}{$p(\mathcal{X}^{\mathrm{dom}},\mathcal{Y}^{\mathrm{dom}})$}. Without adapting to any specific target domain data \textcolor{black}{ $\mathcal{D}^{\mathrm{tar}}=\{(\mathbf{X}_i^{\mathrm{tar}},y_i^{\mathrm{tar}})\}_{i=1}^{N_{\mathrm{tar}}}$} (where $N_{\mathrm{tar}}$ denotes the number of samples of the target domain), the goal in DG is to generalize over unseen target domain distributions \textcolor{black}{$p(\mathcal{X}^{\mathrm{tar}},\mathcal{Y}^{\mathrm{tar}})$}, where \textcolor{black}{$\mathrm{tar}\in[K+1,\infty)$}~\citep{gulrajani2020search}. In the following sub-sections, we have discussed the different DG approaches to address the cross-corpora LID problem.

% \vspace{-0.15cm}
\subsection{Domain adversarial ECAPA-TDNN architecture}
\label{sec:6_c}
% \vspace{-0.15cm}

\begin{figure*}[!t]
% \vspace{-.6cm}
    \centering
    \includegraphics[trim=.01cm .01cm .01cm 
    .01cm,clip,width=\textwidth]{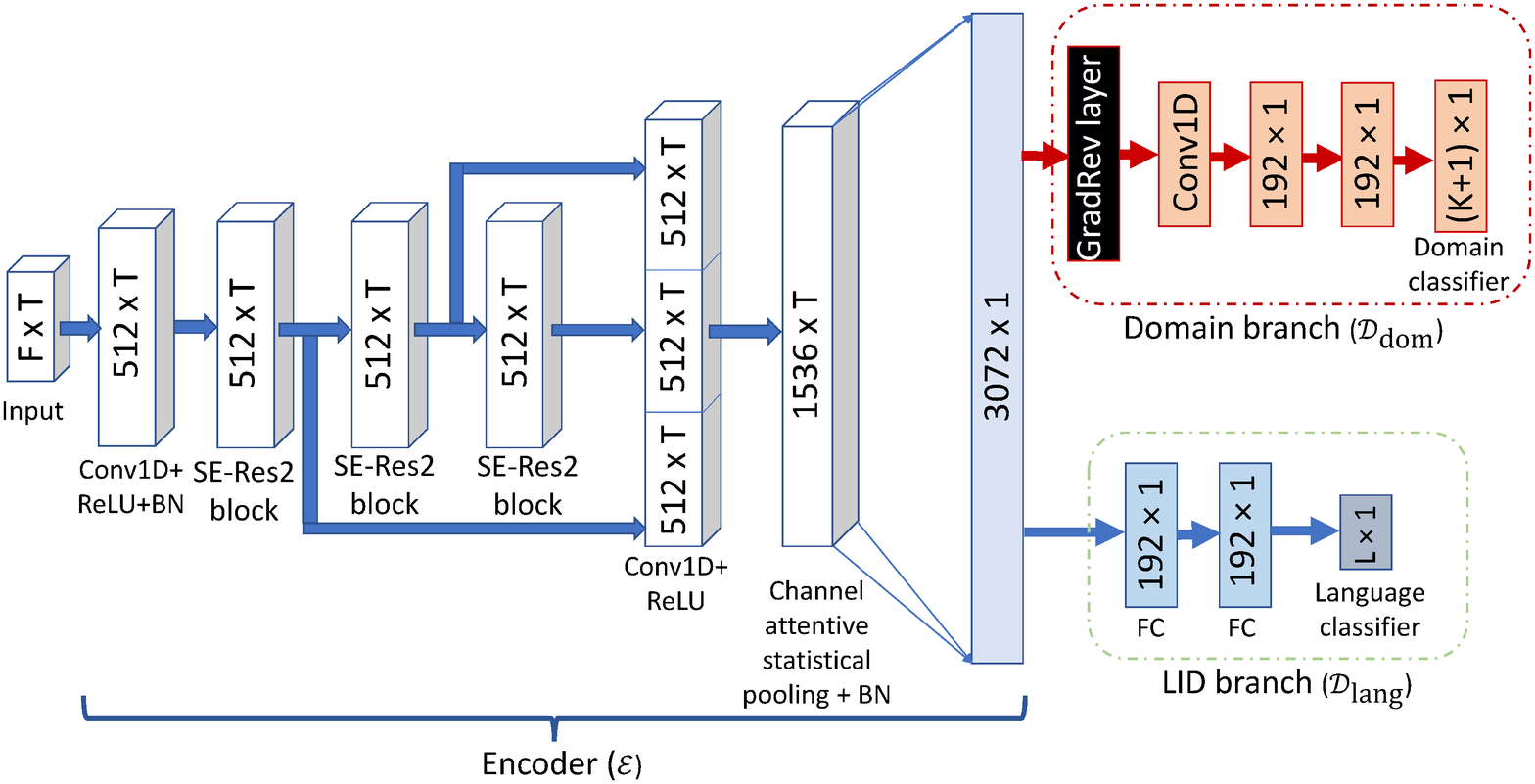}
    %\vspace{-.3cm}
    %\caption{Adversarial ECAPA-TDNN architecture with additional domain-invariant branch (parallel to the LID branch) after the pooling layer.}
    \caption{Adversarial ECAPA-TDNN architecture for domain generalization. \textcolor{black}{Here, $K=7$ (with ``+1" denoting the original non-augmented class) denotes the total number of pseudo-domain corresponding to each signal-level category. $L$ denotes the number of target languages which is five in the present closed-set LID study.}}
    \label{fig:ecapa_adversarial}
    %\vspace{-.56cm}
\end{figure*}

This work further extends the state-of-the-art ECAPA-TDNN architecture with an additional adversarial domain classifier branch for learning domain-invariant language information. As shown in Fig.~\ref{fig:ecapa_adversarial}, the proposed architecture has an encoder ($\mathcal{E}$) which is composed of the frame-level SE-Res2 blocks followed by the attentive pooling layer. For each utterance, the encoder maps the input feature vector $\mathbf{X}$ to $Z$. After the encoder, we have used two segment-level classification branches; LID branch (\textcolor{black}{$\mathcal{D}_\mathrm{lang}$}) and domain branch (\textcolor{black}{$\mathcal{D}_\mathrm{dom}$}). $\mathcal{D}_\mathrm{lang}$ is composed of two fully connected layers. For each $i$-th utterance, it outputs the language posterior probabilities $p(\mathbf{y}|\mathbf{X}_i;\theta_E, \theta_L)$. Here, $\theta_E$ and $\theta_L$ denote the learnable parameters of the encoder and LID branch, respectively. The additional domain classifier branch $\mathcal{D}_\mathrm{dom}$ (with learnable parameters denoted as $\theta_D$) predicts the pseudo-domain posterior probabilities \textcolor{black}{$p(\mathbf{y}^{\mathrm{dom}}|\mathbf{X}_i;\theta_E, \theta_D)$} as a $K+1$ class softmax vector with \textcolor{black}{$K$ signal-level augmentation categories and the additional class of original non-augmented utterances. Here, we do not include the sub-category classes in the domain classifier because all of them perturb the audio in a similar manner. Further, too many similar output classes due to the sub-categories can make the domain classification task challenging which can hinder the main LID task.} Here, \textcolor{black}{$\mathcal{Y}^{\mathrm{dom}} \in \{0, 1, \ldots, K\}$} denotes the pseudo-domain label space. The $\mathcal{D}_\mathrm{dom}$ branch consists of a Conv1D layer followed by two fully connected layers. It is connected to the $\mathcal{E}$ module by a \emph{gradient reversal layer} (GradRev)~\citep{ganin2016domain} that flips the sign of the gradient flows into the encoder during the backpropagation. During training, the pseudo-domain data \textcolor{black}{$\{(\mathbf{X}^{\mathrm{dom}}, \mathbf{y}, \mathbf{y}^{\mathrm{dom}})\}\:\forall \; {\mathrm{dom}} \in [0,K]$} is fed as input. We train the model by optimizing the combined adversarial loss function ($\mathcal{L}_{\mathrm{adv}}$) defined as

%\vspace{-0.1cm}
\begin{equation}
    \begin{array}{r}
    \mathcal{L}_{adv}(\theta_E,\theta_L,\theta_D) = \sum\limits_{i}[(1-\lambda_{\mathrm{adv}})\mathcal{L}_{\mathrm{lang}}(y_i,\mathcal{D}_\mathrm{lang}(\mathcal{E}(\mathbf{X}_i));\theta_E,\theta_L)\\ - \lambda_{\mathrm{adv}} \; \mathcal{L}_{\mathrm{dom}}(\mathcal{D}_\mathrm{dom}(y^{\mathrm{dom}}_i,\mathcal{E}(\mathbf{X}_i));\theta_E,\theta_D)]. \end{array}
    \label{eq:adv_loss}
\end{equation}

Here, $i$ denotes index of individual utterances, $\mathbf{X}_i \in \mathcal{X}$. The total loss ($\mathcal{L}_{\mathrm{adv}}$) is composed by the language classification loss $\mathcal{L}_{\mathrm{lang}}$  and domain classification loss $\mathcal{L}_{\mathrm{dom}}$. The parameter $\lambda_{\mathrm{adv}} \in [0,1]$ decides the weights of the two losses in the total loss. By optimizing the min-max problem, $\mathcal{L}_{\mathrm{lang}}$ trains $\theta_L$ and $\theta_E$ to improve the LID accuracy. Simultaneously, $\theta_D$ is estimated to improve the pseudo-domain recognition performance. The gradient reversal layer flips the sign of $\delta \mathcal{L}_{\mathrm{dom}}/ \delta \theta_D$, resulting in $\theta_E$ to learn domain-invariant language discriminative information.

\subsection{Multitask learning for cross-corpora generalization}
\label{sec:6_d}
We have explored multitask learning (MTL)~\citep{zhang2021survey,ruder2017overview,caruana1997multitask} in the cross-corpora LID framework with the aim of learning pseudo-domain aware representations that can improve generalization. \textcolor{black}{Joint learning of multiple features or multiple tasks has been shown to improve the robustness for cross-channel LID test samples~\citep{li2021deep}.} Using MTL, our LID model jointly learns an additional eight-class domain classification task~\citep{zhou2022domain}. Thus, the LID is expected to capture the domain-specific language discriminative information generalized across multiple pseudo-domains. In this experiment, the extension to the ECAPA-TDNN architecture is similar to that of the adversarial scenario (see Section~\ref{sec:6_c}) except for the removal of the gradient reversal layer~\citep{adi2019reverse}. We can express the total loss to be optimized as

%\vspace{-0.1cm}
\begin{equation}
    \begin{array}{r}
    \mathcal{L}_{\mathrm{MTL}}(\theta_E,\theta_L,\theta_D) = \sum\limits_{i}[\mathcal{L}_{\mathrm{lang}}(y_i,\mathcal{D}_\mathrm{lang}(\mathcal{E}(\mathbf{X}_i));\theta_E,\theta_L)\\ + \lambda_{\mathrm{MTL}} \; \mathcal{L}_{\mathrm{dom}}(y^{\mathrm{dom}}_i,\mathcal{D}_{\mathrm{dom}}(\mathcal{E}(\mathbf{X}_i));\theta_E,\theta_D)]. \end{array}
    \label{eq:self_loss}
\end{equation}

In Eq.~\ref{eq:adv_loss} and Eq.~\ref{eq:self_loss}, the major difference lies in the sign of the $\mathcal{L}_{\mathrm{dom}}$ component. So, basically, these two experiments attempt to solve the domain-mismatch problem following two opposite strategies: domain-invariant and domain-aware~\citep{adi2019reverse}. In domain generalization, we do not have information about any specific target domain. Therefore, both of these opposite but complementary DG approaches can be suitable across the different unseen domains that may be encountered in real-world applications.
\subsection{MMD-based domain generalization approaches}
The most fundamental assumption in machine learning is that the training and evaluation data are independent and identically distributed (i.i.d.). However, this assumption is often violated due to the domain mismatch. For improving generalization, utilizing the pseudo-domains, our idea is to train the LID model in such a way that the distribution mismatch among the training conditions is reduced. In this work, we explore \emph{maximum mean discrepancy} (MMD)~\citep{gretton2006kernel} based techniques for domain generalization. We investigate MMD for two main reasons. First, this efficiently minimizes the distribution shift between source and target data~\citep{gretton2006kernel}. Second, the language recognition literature successfully used MMD for reducing domain mismatch~\citep{duroselle20_interspeech}.

MMD is the metric for measuring the distance between two distributions~\citep{gretton2006kernel}. In DA literature, MMD-based loss ($\mathcal{L}_{\mathrm{MMD}}$) is optimized to minimize the statistical moments differences between the source and target domains~\citep{zhu2020deep}. In this work, we first utilize the pseudo-domains as a target domain so that the MMD-based DA technique is applied under the DG configurations. Thereafter, we extend the MMD-based DG model by proposing multi-domain extension. For the conventional MMD-based DG experiments, we have treated the non-augmented training utterances as source domain \textcolor{black}{($\mathcal{S}^\mathrm{src}$)} and all the pseudo-domains \textcolor{black}{($\mathcal{S}^\mathrm{dom}$)} as a target domain.

Let the feature space distribution of the source and target domain be denoted as \textcolor{black}{$p^{\mathrm{src}} =  p(\mathcal{X}^{\mathrm{src}})$ and $p^{\mathrm{dom}} =  p(\mathcal{X}^{\mathrm{dom}})$}, respectively. We first compute the MMD-loss as the squared difference between the means of the source and target representations mapped to a \emph{reproducing kernel Hilbert space} (RKHS)~\citep{gretton2006kernel} and this is defined as

\begin{equation}
    d_\mathcal{H} = \left|\left| \sum_{\mathbf{X}^{\mathrm{src}}_i \in \mathcal{S}^{\mathrm{src}}} \phi(\mathbf{X}^{\mathrm{src}}_i) \;-\; \sum_{\mathbf{X}^{\mathrm{dom}}_i \in \mathcal{S}^{\mathrm{dom}}} \phi(\mathbf{X}^{\mathrm{dom}}_i)\right|\right|_{\mathcal{H}}^2.
    \label{eq:mmd1}
\end{equation}

In Eq.~\ref{eq:mmd1}, $i$ 
 is the utterance indices, $\mathbf{X}^{\mathrm{src}} \in \mathcal{X}^{\mathrm{src}}$, $\mathbf{X}^{\mathrm{dom}} \in \mathcal{X}^{\mathrm{dom}}$. $\mathcal{H}$ denotes the RKHS space corresponding to a positive semi-definite kernel \textcolor{black}{$\mathcal{K}$}. The MMD-loss is jointly optimized with the LID loss to make the LID model robust against the domain shifts. For practical purposes, an empirical estimate of the MMD-loss is computed by Eq.~\ref{eq:mmd2} assuming a positive definite kernel $\mathcal{K}(\mathbf{X}^{\mathrm{src}},\mathbf{X}^{\mathrm{dom}}) = \langle\phi(\mathbf{X}^{\mathrm{src}}),\phi(\mathbf{X}^{\mathrm{dom}})\rangle$. Here, $\phi$ maps the inputs from feature space to the RKHS, and $\langle~,~\rangle$ operator denotes the inner product.
% \vspace{-0.1cm}
\begin{equation}
        \hat{d}_\mathcal{H} = \left|\left|\frac{1}{N_{\mathrm{src}}} \sum_{i=1}^{N_{\mathrm{src}}} \phi(\mathbf{X}^{\mathrm{src}}_i) \;-\; \frac{1}{N_{\mathrm{dom}}} \sum_{i=1}^{N_{\mathrm{dom}}} \phi(\mathbf{X}^{\mathrm{dom}}_i)\right|\right|_{\mathcal{H}}^2
        \label{eq:mmd2}
\end{equation}

Here, $N_{\mathrm{src}}$ and $N_{\mathrm{src}}$ denote the number of source and target domain utterances, respectively. Further simplification of Eq.~\ref{eq:mmd2} leads to:

\begin{equation}
%\vspace{-0.15cm}
\begin{array}{c}
\hat{d}_\mathcal{H} = \frac{1}{N_{\mathrm{src}}^2}\sum\limits_{i=1}^{N_{\mathrm{src}}}\sum\limits_{j=1}^{N_{\mathrm{src}}} \phi(\mathbf{X}^{\mathrm{src}}_i)\phi(\mathbf{X}^{\mathrm{src}}_j) + \frac{1}{N_{\mathrm{dom}}^2}\sum\limits_{i=1}^{N_{\mathrm{dom}}}\sum\limits_{j=1}^{N_{\mathrm{dom}}} \phi(\mathbf{X}^{\mathrm{dom}}_i)\phi(\mathbf{X}^{\mathrm{dom}}_j) - \frac{2}{N_{\mathrm{src}} N_{\mathrm{dom}}}\sum\limits_{i=1}^{N_{\mathrm{src}}}\sum\limits_{j=1}^{N_{\mathrm{dom}}} \phi(\mathbf{X}^{\mathrm{src}}_i)\phi(\mathbf{X}^{\mathrm{dom}}_j)
\end{array}
\label{eq:mmd3}
% \vspace{-.5cm}
\end{equation}

Using multiple-kernel extensions in MMD-loss enhances the classification performance~\citep{long2015learning}. In this work, we use kernel $\mathcal{K}$ as the combination of $M$ radial basis function (RBF) with different variances:
% \vspace{-0.05cm}
\begin{equation}
    \mathcal{K}(\mathbf{X}_i,\mathbf{X}_j) = \sum\limits_{m=1}^{M} exp\left(\frac{ (\mathbf{X}_i - \mathbf{X}_j)^T(\mathbf{X}_i-\mathbf{X}_j)}{\sigma^2_m}\right)
    \label{eq:mmd4}
    %\vspace{-.1cm}
\end{equation}
Where, 
\begin{equation}
% \vspace{-0.05cm}
    \sigma^2_m = \frac{2^m}{N_{\mathrm{src}} N_{\mathrm{dom}}}\sum\limits_{i=1}^{N_{\mathrm{src}}}\sum\limits_{j=1}^{N_{\mathrm{dom}}}(\mathbf{X}_i - \mathbf{X}_j)^{\top}(\mathbf{X}_i-\mathbf{X}_j)
    \label{eq:mmd5}
    % \vspace{-.1cm}
\end{equation}

The combined loss is defined as
%\vspace{-0.1cm}
\begin{equation}
\mathcal{L}_{\mathrm{mmd}-\mathrm{tot}} =  \mathcal{L}_{\mathrm{lang}} + \lambda_{\mathrm{mmd}} \: \hat{d}_\mathcal{H}.
\label{eq:mmd_loss}
%\vspace{-.1cm}
\end{equation}

\begin{figure*}[!t]
% \vspace{-0.1cm}
    \centering
    \includegraphics[trim=0cm 0cm 0cm 0cm,clip,width=\linewidth]{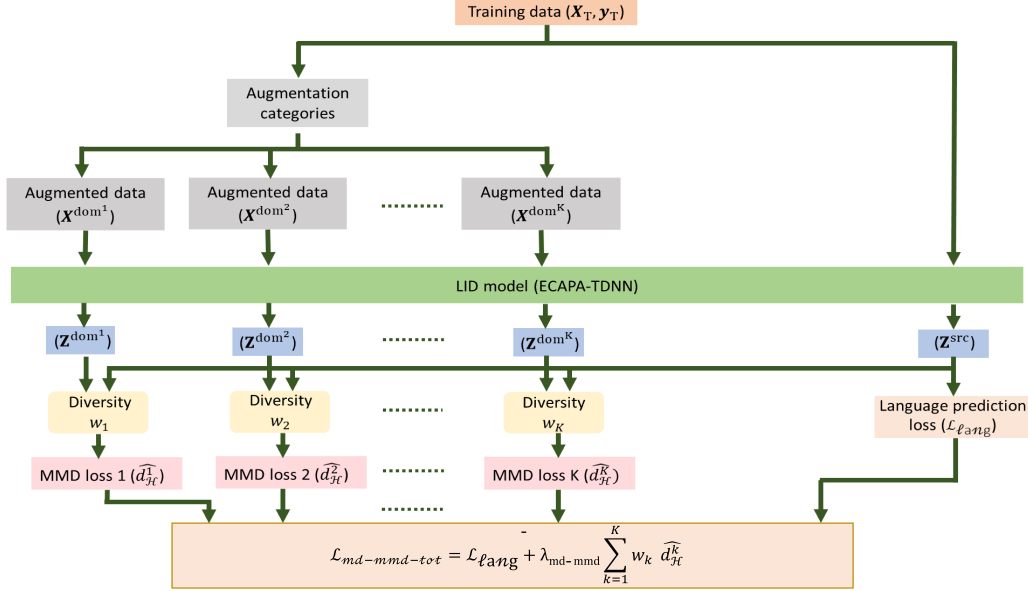}
    %\vspace{-.2cm}
    \caption{Working principle of the MD-MMD based ECAPA-TDNN model with $\mathcal{L}_{\mathrm{lang}}$ denoting the AM-softmax language prediction loss and $\hat{d}^k_\mathcal{H}$ denoting individual multi-domain MMD loss components.}
    \label{fig:MD-MMD}
\vspace{-.45cm}
\end{figure*}

% The joint optimization makes the ECAPA-TDNN learn the language classes while producing similar representation of the source and target domain inputs. The conventional MMD-based optimization approaches are suitable for any specific target domain. 
In Eq.~\ref{eq:mmd_loss}, $\lambda_{\mathrm{mmd}}$ is the scalar to determine the relative weights of the LID loss and MMD loss. The joint optimization learns the language-discriminating representations while producing a more similar representation of the source and target domain.

In Eq.~\ref{eq:mmd_loss}, all the pseudo-domains are collectively treated as a single target domain $\mathcal{S}^{\mathrm{dom}}$ with the single MMD-loss component $\hat{d}_\mathcal{H}$. To utilize the finer domain differences among the pseudo-domains more efficiently, we propose to extend the conventional MMD approach into \emph{multi-domain MMD} (MD-MMD) loss. The proposed MMD method efficiently incorporates the different sources of diversity for each pseudo-domain. For each pseudo-domain \textcolor{black}{$\mathcal{S}^{\mathrm{dom}^k}$}, where $k=[1:K]$, we compute separate MMD loss component $\hat{d}^k_\mathcal{H}$:

\begin{equation}
% \vspace{-0.15cm}
        \hat{d}^k_\mathcal{H} =  \left|\left|\frac{1}{N_{\mathrm{src}}} \sum_{i=1}^{N_{\mathrm{src}}} \phi(\mathbf{X}^{\mathrm{src}}_i) \;-\; \frac{1}{{N_{\mathrm{dom}}^k}} \sum_{i=1}^{{N_{\mathrm{dom}}^k}} \phi(\mathbf{X}^{\mathrm{dom}^k}_i)\right|\right|_{\mathcal{H}}^2
        \label{eq:mmd2k}
\end{equation}

Here, \textcolor{black}{$N_{\mathrm{dom}^k}$} denotes the number of training utterances for $\mathcal{S}^{\mathrm{dom}^k}$. Each component is then optimized simultaneously with $\mathcal{L}_{\mathrm{lang}}$. The total loss function for the MD-MMD is:

% \vspace{-0.15cm}
\begin{equation}
\mathcal{L}_{\mathrm{\mathrm{md}-\mathrm{mmd}-\mathrm{tot}}} =  \mathcal{L}_{\mathrm{lang}} + \lambda_{\mathrm{md-mmd}} \sum\limits_{k=1}^{K} w_k \: \hat{d}^k_\mathcal{H}
\label{eq:md-mmd_loss}
\end{equation}

In Eq.~\ref{eq:md-mmd_loss}, the $w_k$ are the relative weights for each pseudo-domains in the MD-MMD loss. One simple way is to set $w_k = 1/K\; \forall \; k \in [1,K]$. However, different pseudo-domains offer different levels of domain diversity in the source domain. Therefore, the domain with the most diversity measure from the source domain can be weighted the most. In the proposed approach, $w_k$ are set according to the corresponding normalized domain diversity measures (see Section~\ref{sec:6_c}) from the source domain:

\begin{equation}
w_k = \left[\mathbf{D}_{KL}^{\{0,k\}}/\sum\limits_{k=1}^{K} \mathbf{D}_{KL}^{\{0,k\}}\right].    
\end{equation}

The detailed working principle of the proposed MD-MMD model is illustrated in Fig.~\ref{fig:MD-MMD}. 

%\vspace{-0.15cm}
\section{Results \& analysis}
\label{Sec:7}
%\vspace{-0.15cm}
In this section, we present and analyze the LID performance results of the different experiments that we conduct in this study. Following the NIST LRE~\citep{sadjadi20182017} and OLR challenges~\citep{li2020ap20}, the LID performances are reported using $EER~(\%)$ / $C_{\mathrm{avg}} * 100$ (in the Tables denoted as $EER / C_{\mathrm{avg}}$). \textcolor{black}{We primarily focus on the mismatch between the cross-corpora and same-corpora performance metrics during the evaluations.} For the proposed approaches, compared to the baseline, the relative improvements in the same-corpora LID performances are also studied. We first compare the LID performances of the x-vector baseline TDNN and the ECAPA-TDNN architecture. Then, we individually assess the LID performance of each augmentation category. Following that, we train and evaluate the different domain diversification approaches discussed earlier. Finally, we utilize the diversified data as pseudo-domains and evaluate the LID performances of the DG extensions.

%\vspace{-0.1cm}
\subsection{Baseline LID performances}
%\vspace{-0.05cm}
We first train LID models using the baseline x-vector TDNN and ECAPA-TDNN architectures. The corresponding same-corpora and cross-corpora evaluation results are presented in~Table~\ref{tab:tdnn_eval}. 
\begin{table*}
\centering
\caption{LID performance evaluation of the TDNN architectures in $EER~(\%)$ / $C_{\mathrm{avg}} * 100$.}
\label{tab:tdnn_eval}
\resizebox{\linewidth}{!}{%
\begin{tabular}{|c|c|c|c|c|c|c|} 
\hline
\multirow{3}{*}{\begin{tabular}[c]{@{}c@{}}Training \\corpus\end{tabular}} & \multicolumn{3}{c|}{Baseline TDNN} & \multicolumn{3}{c|}{ECAPA-TDNN} \\ 
\cline{2-7}
 & IIITH-test & LDC-test & KGP-test & IIITH-test & LDC-test & KGP-test \\ 
\cline{2-7}
 & EER / $C_\mathrm{avg}$ & EER / $C_\mathrm{avg}$ & EER / $C_\mathrm{avg}$ & EER / $C_\mathrm{avg}$ & EER / $C_\mathrm{avg}$ & EER / $C_\mathrm{avg}$ \\ 
\hline
IIITH & 12.40 / 13.82 & 43.10 / 46.23 & 30.30 / 29.91 & \textbf{9.74} / \textbf{11.51} & 42.43 / 44.82 & 34.83 / 32.62 \\ 
\hline
LDC & 50.08 / 47.34 & 22.95 / 26.40 & 45.60 / 45.92 & 46.35 / 43.21 & \textbf{21.86} / \textbf{25.70} & 42.95 / 39.59 \\ 
\hline
KGP & 37.55 / 34.91 & 51.80 / 46.72 & \textbf{11.18} / 10.82 & 36.52 / 33.59 & 47.25 / 45.49 & 12.36 / \textbf{8.75} \\
\hline
\end{tabular}
}
\end{table*}
The results demonstrate that ECAPA-TDNN outperforms the x-vector TDNN for most of the evaluation cases. \textcolor{black}{We observe that only for some cases during the evaluation with KGP-test data, x-vector shows slightly improved performances. The additional modifications that are modeled in the ECAPA-TDNN compared to the x-vector potentially degrade the robustness towards the non-lingual mismatches contained in the KGP-test data. KGP-test subset contains only 60 chunks (each of 3~s duration) from each language. Whereas IIITH-test contains a total of 4878 numbers of 3~s chunks, and the LDC-test contains a total of 24617 3~s chunks. Due to the extremely small size of the KGP-test, there can be chances of sample biases that lead to inferior ECAPA-TDNN performance.} Nevertheless, since ECAPA-TDNN achieves better LID performance for the majority of the LID evaluations, we only consider the ECAPA-TDNN based LID models for the subsequent experiments. \textcolor{black}{For both architectures, we observe a prominent performance mismatch between the same-corpora and cross-corpora evaluations. The cross-corpora LID experiments achieve EER of more than $30\%$, indicating unreliable language predictions. The higher EER also reflects that achieving decent generalization over different LID corpora is a challenging task.} %However, the key focus of our work is to examine the relative change of cross-corpora performances (compared to the performances reported in Table~\ref{tab:tdnn_eval}) for the subsequent experiments.}

\subsection{LID performances: Individual augmentation categories}

\begin{table*}
\centering
\caption{LID performances (in $EER (\%)$ / $C_{\mathrm{avg}}*100$) due to individual signal-level augmentation category in the IIITH training data.}
\label{tab:aug_cross_corpora_LID}
\resizebox{\linewidth}{!}{%
\begin{tabular}{|l|l|l|l|l|l|l|l|l|} 
\hline
\begin{tabular}[c]{@{}l@{}}Testing\\corpus\end{tabular} & \begin{tabular}[c]{@{}l@{}}Without\\augmentation~\end{tabular} & \begin{tabular}[c]{@{}l@{}}Non-speech\\addition~($A^{1}$)\end{tabular} & \begin{tabular}[c]{@{}l@{}}Signal parameter\\perturbation~($A^{2}$)\end{tabular} & \begin{tabular}[c]{@{}l@{}}Bandwidth\\augmentation~($A^{3}$)\end{tabular} & \begin{tabular}[c]{@{}l@{}}Environmental\\variations~($A^{4}$)\end{tabular} & \begin{tabular}[c]{@{}l@{}}Speech \\enhancements~($A^{5}$)\end{tabular} & \begin{tabular}[c]{@{}l@{}}Audio\\encoding~($A^{6}$)\end{tabular} & \begin{tabular}[c]{@{}l@{}}Lossy codec\\restoration~($A^{7}$)\end{tabular}  \\ 
\hline
IIITH                                                   & 9.74 / 11.51                                                   & \textbf{8.43} / \textbf{8.73}                                                 & 10.12 / 11.76                                                                       & 9.76 / 11.56                                                                 & 9.48 / 9.46                                                                   & 9.21 / 9.06                                                               & 8.93 / 9.01                                                         & 9.63 / 11.42                                                                  \\
LDC                                                     & 42.43 / 44.82                                                  & 40.33 / 36.78                                                                 & 41.15 / 43.64                                                                       & 43.77 / 46.11                                                                & 39.62 / 37.80                                                                 & 39.84 / 38.32                                                             & \textbf{39.32} / \textbf{36.44}                                     & 42.71 / 44.70                                                                 \\
KGP                                                     & 34.83 / 32.62                                                  & 33.23 / 39.01                                                                 & 39.87 / 38.70                                                                       & 42.06 / 37.32                                                                & 37.08 / 33.72                                                                 & 35.06 / 31.91                                                             & \textbf{32.36} / \textbf{29.98}                                     & 38.26 / 35.59                                                                 \\
\hline
\end{tabular}
}
\end{table*}

For each of the seven signal-level augmentation categories $A^{k}$, for each corresponding sub-category, we create an augmented copy of the training utterances. Thereafter, for each augmentation category, we randomly sample (with fold-factor $\gamma = 1$) the augmented utterances and combine them with the original non-augmented training data ($A^{0}$). We thus create seven sets of combined augmented audio data and train individual LID models. Table~\ref{tab:aug_cross_corpora_LID} shows the LID performance comparison for each of the signal-level augmentation categories. For comparative assessments, the LID performance of the ECAPA-TDNN models, trained with the non-augmented original training data, are also presented in Table~\ref{tab:aug_cross_corpora_LID}.
 
For all three training corpora, we observe LID performance improvements in similar ranges due to the explored augmentation categories. Hence, in Table~\ref{tab:aug_cross_corpora_LID} and in the following domain diversification experiments, we have presented the LID performances with only the IIITH-trained scenario. From Table~\ref{tab:aug_cross_corpora_LID}, we observe that for same-corpora evaluation, the non-speech addition category performs the best. For both LDC and KGP cross-corpora evaluations, codec-encoding augmentation shows the best performance improvements. The proposed speech-enhancement based augmentation and environmental augmentations also exhibit consistently promising performance improvements for both the cross-corpora evaluations. We then try to associate these cross-corpora evaluation observations with the available meta information. Contrary to the broadcast news studio recordings of the IIITH and KGP, the LDC data contains telephonic conversations, recorded in both indoor and outdoor environments. So, the diverse background noise related mismatches can be suppressed using speech enhancement based augmentations. The environmental augmentations helps in reducing corpora-mismatch in terms of recording room environments. LID performance improvements due to the codec encoding augmentation can be justified by the fact that the intermediate data processing and storage protocols of the different corpora may be different.

\subsection{LID performance of different augmentation fold-factors}
\label{sec:fold}
%\vspace{-0.05cm}
We experimentally optimize the augmentation fold-factor ($\gamma$) that controls the ratio of the augmented and original training utterances in the combined augmented training set. We have trained independent LID models using $\gamma = \{0.5, 1, 2, 3, 4\}$ and compared their LID performances in Table~\ref{tab:aug_folds}. The results indicate that as $\gamma$ increases from 0.5 to 3, in general, both same-corpora and cross-corpora LID performance improve. \textcolor{black}{For IIITH-trained LID model, best same-corpora EER of $7.68 \%$ is achieved for $\gamma=3$. For the same train-test pair, the lowest $C_\mathrm{avg}$ of $9. 14$ is attained for $\gamma=2$. For the cross-corpora evaluations with LDC-test, we observe that the EER attains the lowest value at $\gamma=2$. Whereas, the lowest $C_{\mathrm{avg}}$ for KGP-test is achieved for $\gamma=3$}. This echos the fact that the performances of the TDNN-based models improve with more training data incurred due to augmentation~\citep{snyder2018x}. But for almost all the cases, the LID performances start to degrade beyond $\gamma = 3$. This is an interesting observation that can be justified by saying that beyond $\gamma=3$, there are too many acoustically diversified training samples compared to the original training data. So, the LID models may start learning the non-lingual diversities instead of the actual language-discriminating cues present in the original data~\citep{sturm2014simple}. Moreover, a smaller value of the fold-factor, $\gamma = 2, 3$, also reduces computation and the model training time.

\textcolor{black}{For the subsequent experiments, we use $\gamma=2$ to sample utterances from different signal-level augmentation categories. 
This sampled augmentation set is denoted as “signal-level ($\gamma = 2$)” in Table~\ref{tab:aug_synthetic} and “Augmentation ($\gamma = 2$)” in Table~\ref{tab:adversarial_LID_results}.} Sampled augmentation sets with $\gamma=2\; \mbox{or} \;3$ is optimum in terms LID performances in our work. Further, for $\gamma=2$, we also need a lesser computational cost. \textcolor{black}{It is an interesting future work to verify the optimum augmentation fold-factors across the different speech-based classification tasks.}

\begin{table}
\centering
\caption{Impact of augmentation fold-factors on LID performance (in EER ($\%$) / $C_{\mathrm{avg}}*100$) using the IIITH as training corpus.}
\label{tab:aug_folds}
\resizebox{.65\linewidth}{!}{%
\begin{tabular}{|c|c|c|c|} 
\hline
\multirow{2}{*}{Fold-factor ($\gamma$)} & IIITH-test & LDC-test & KGP-test \\ 
\cline{2-4}
 & EER / $C_\mathrm{avg}$ & EER / $C_\mathrm{avg}$ & EER / $C_\mathrm{avg}$ \\ 
\hline
0.5 & 9.18 / 10.92 & 39.43 / 43.79 & 37.56 / 34.89 \\ 
\hline
1 & 8.75 / 10.01 & 39.61 / 43.96 & 38.65 / 36.70 \\ 
\hline
2 & 8.00 / \textbf{9.14} & \textbf{38.19} / 42.71 & 38.88 / 35.52 \\ 
\hline
3 & \textbf{7.68} / 9.28 & 38.64 / \textbf{42.40} & 36.80 / \textbf{32.92} \\ 
\hline
4 & 8.03 / 9.68 & 38.89 / 43.20 & \textbf{35.73} / 34.30 \\
\hline
\end{tabular}
}
\end{table}
%\vspace{-0.1cm}

\subsection{LID performance of signal-feature level cascaded augmentations}
% \vspace{-0.1cm}

We apply cascaded data augmentation by performing feature-level augmentations during the feature extraction steps of the signal-level augmented utterances. We choose all seven signal-level augmentation categories and sample them with $\gamma = 2$. For feature-level augmentation, we evaluate SpecAug and mixup separately. The corresponding LID performances (for IIITH as training data) are presented in Table~\ref{tab:mixup}.

\begin{table}[!t]
\centering
\caption{LID performances (in EER ($\%$) / $C_{\mathrm{avg}}*100$) by cascading the signal-level and feature-level augmentations using the IIITH as training corpus.}
\label{tab:mixup}
\resizebox{.8\linewidth}{!}{%
\begin{tabular}{|c|c|c|c|} 
\hline
\multirow{2}{*}{Augmentation} & IIITH-test & LDC-test & KGP-test \\ 
\cline{2-4}
 & EER / $C_\mathrm{avg}$ & EER / $C_\mathrm{avg}$ & EER / $C_\mathrm{avg}$ \\ 
\hline
Signal-level ($\gamma =2 $) & \textbf{8.00} / \textbf{9.14} & \textbf{38.19} / 42.71 & 38.88 / 35.52 \\ 
\hline
Signal-level followed by SpecAug & 9.82 / 11.20 & 48.67 / 45.16 & 39.74 / 40.24 \\ 
\hline
Signal-level followed by Mixup & 9.26 / 10.93 & 39.70 / \textbf{40.61} & 45.01 / 38.94 \\ 
\hline
Diversity-aware cascading & 9.13 / 10.74 & 39.70 / 43.16 & \textbf{35.17} / \textbf{29.32} \\ 
\hline
Signal-level followed by cascading & 8.61 / 10.22 & 39.16 / 43.04 & 37.75 / 35.84 \\
\hline
\end{tabular}
}
\end{table}

\textcolor{black}{Analyzing Table~\ref{tab:mixup}, we observe that cascading the different signal-level and feature-level augmentation categories does not help in improving the LID generalization. It may be possible that the signal-feature augmented utterances are over-diversified, and the language classification task becomes challenging. As a consequence, in Section~\ref{sec:7_e}, we have reported the experiments of the diversity-aware cascaded augmentations by considering the multiple signal-level augmentation categories.}

%\vspace{-0.1cm}
\subsection{LID performance: Diversity-aware cascaded augmentations}
\label{sec:7_e}
%\vspace{-0.1cm}

To augment the utterances in a more realistic manner, we propose cascading different signal-level augmentation categories to create further diversified training utterances. For effective diversification, we find out the pair-wise maximally diversified augmentation categories for cascading.

\subsubsection{Metrics for measuring the domain diversity}

For each signal-level augmentation category $A^k$, embeddings ($Z^{k}$) are extracted for the corresponding utterances. For embedding extraction, we use the LID model trained on the combined augmentation IIITH training set (with $\gamma=2$). We also extract embeddings for the original non-augmented IIITH utterances and denote them as $Z^0$. Next, we measure the embedding distances between $Z^{0}$ and $Z^{k}\; \forall\; k \in[1,7]$ using three different metrics, cosine distance ($\mathbf{D}_{cos}$), KL-divergence ($\mathbf{D}_{KL}$), and symmetric KL divergence ($\mathbf{D}_{KL-s}$). For each metric, we have correlated the embedding distance with the corresponding validation  $C_{\mathrm{avg}}$ (primary evaluation metric) scores and presented the results in Table~\ref{tab:correlate}. As the embedding distances are not guaranteed to follow Gaussian distribution, we use the \emph{Spearman correlation coefficients} for the correlation computation. We have finally selected the KL divergence for measuring the domain diversity distance because it shows the most negative correlation with the $C_{\mathrm{avg}}$ scores. The more the KL divergence between the original data and an augmentation category, the more the diversity expected, and eventually the lesser the corresponding $C_{\mathrm{avg}}$. Further, the non-commutative characteristics (not present in $\mathbf{D}_{KL-s}$) of the KL divergence also support the basic assumption of the cascaded augmentations: $A^{\{i,j\}} \neq A^{\{j,i\}}$.

\begin{table}[!htbp]
%\vspace{-.1cm}
\centering
\caption{Different domain diversity measures and their correlations with $C_{\mathrm{avg}}$ on IIITH validation data.}
%\vspace{-0.1cm}
\label{tab:correlate}
\begin{footnotesize}
\resizebox{.8\linewidth}{!}{%
\begin{tabular}{|lllllllll|} 
\hline
                  & $A^{1}$               & $A^{2}$                & $A^{3}$                & $A^{4}$               & $A^{5}$               & $A^{6}$               & $A^{7}$                & Correlation  \\ 
\hline
$\mathbf{D}_{cos}$  & 0.127                 & 0.106                  & 0.221                  & 0.123                 & 0.122                 & 0.009                 & 0.008                  & -0.107       \\
$\mathbf{D}_{KL}$   & 0.353                 & 0.133                  & 0.257                  & 0.188                 & 0.205                 & 0.171                 & 0.065                  & -0.428       \\
$\mathbf{D}_{KL-s}$ & 0.303                 & 0.177                  & 0.252                  & 0.211                 & 0.232                 & 0.204                 & 0.128                  & -0.428       \\
$C_{\mathrm{avg}}$           & 7.11 & 11.30 & 11.07 & 7.82 & 7.98 & 7.20 & 9.95 & -            \\
\hline
\end{tabular}
}
\end{footnotesize}
%\vspace{-0.3cm}
\end{table}

\subsubsection{LID performance: Maximally diversity-aware cascaded augmentation}

Next, we present the LID performance for the proposed cascaded augmentation method. The aim of this experiment is to increase the diversity in the training data more realistically by cascading signal-level augmentation categories, which are mutually the most diverse. Considering all the $\mathbf{D}_{KL}^{\{i,j\}}$ values, for each $A^{i}$, we cascade its utterances with another augmentation category $A^{j}$ such that $\mathbf{D}_{KL}^{\{i,j\}}$ maximum. Table~\ref{tab:aug_synthetic} shows the pairwise divergence of all augmentation pairs for the IIITH corpus. For each augmentation category, we create one cascaded augmented set. We then sample and combine these cascaded sets with the original non-augmented utterances by $\gamma=2$ and train the LID model. Further, we train another LID model with the combined utterances of original data, signal-level augmented data ($\gamma = 1$), and cascaded augmented data ($\gamma = 1$).  The last two rows of Table~\ref{tab:mixup} show the LID performance using the maximally diversity-aware cascaded augmented utterances. 

\begin{table}[!htbp]
%\vspace{-.1cm}
\ra{1}
\tiny
\centering
\caption{Pairwise KL divergences between the embeddings of the explored signal-level augmentation categories for the IIITH corpus.}
%\vspace{-0.1cm}
\label{tab:aug_synthetic}
\resizebox{.72\linewidth}{!}{%
\begin{tabular}{|lccccccc|} 
\hline
 & \multicolumn{1}{l}{$A^{1}$}  & \multicolumn{1}{l}{$A^{2}$} & \multicolumn{1}{l}{$A^{3}$} & \multicolumn{1}{l}{$A^{4}$} & \multicolumn{1}{l}{$A^{5}$} & \multicolumn{1}{l}{$A^{6}$} & \multicolumn{1}{l|}{$A^{7}$} \\ 
\hline
$A^{1}$ & 0.00 & 0.311 & 0.385 & 0.458 & \textbf{0.713} & 0.169 & 0.312 \\
$A^{2}$ & \textbf{0.313} & 0.00 & 0.271 & 0.151 & 0.178 & 0.228 & 0.101 \\
$A^{3}$ & 0.518 & 0.332 & 0.00 & 0.210 & \textbf{0.609} & 0.391 & 0.309 \\
$A^{4}$ & \textbf{0.341} & 0.155 & 0.191 & 0.00 & 0.287 & 0.278 & 0.194 \\
$A^{5}$ & 0.529 & 0.173 & \textbf{0.532} & 0.329 & 0.00 & 0.340 & 0.251 \\
$A^{6}$ & 0.187 & 0.191 & 0.296 & 0.311 & \textbf{0.393} & 0.00 & 0.180 \\
$A^{7}$ & 0.257 & 0.104 & \textbf{0.280} & 0.174 & 0.210 & 0.195 & 0.00 \\
\hline
\end{tabular}
}
%\vspace{-.05cm}
\end{table}

The proposed maximally diversity-aware cascaded augmentation shows a promising LID performance. Interestingly, for this cascaded augmentation, the relative improvements of the cross-corpora performance are even more than corresponding same-corpora performance improvements. One can keep on cascading more than two augmentation methods to incorporate more diversity in the training data. However, this will increase computational complexity. Furthermore, incorporating more diversity in the data may distort the language-specific cues in the audio data. 

We have also measured the KL divergence of the cascaded-augmented embeddings with the non-augmented embeddings; $\mathbf{D}_{KL}^{\{0,\mbox{cascaded}\}} = 0.286$. Comparing with the $\mathbf{D}_{KL}^{\{0,k\}}$ values for $k\in [1,7]$ (from Table~\ref{tab:correlate}), we find $\mathbf{D}_{KL}^{\{0,\mbox{cascaded}\}}$ to be higher for six out of the seven augmented categories. In this work, for each $A^{i}$ row of Table~\ref{tab:aug_synthetic}, if the column $A^{p}$ shows the highest value, we cascade the two augmentation categories as $A^{\{i,p\}}$. We could also cascade two augmented categories with the highest overall $\mathbf{D}_{KL}^{\{i,j\}}$ values. However, by doing so, it is not guaranteed that the diversities for all the explored augmentation categories, $A^{k} \; \forall \; k=[1,7]$, is preserved by cascading. The objective of this study is to improve generalization on multiple completely independent corpora. The proposed cascading algorithm, which incorporates diversity by ensuring audio from all the explored augmented categories, could be more beneficial in such unseen evaluation scenarios.

%\vspace{-0.1cm}
\subsection{LID performance: Domain generalization perspectives}
% \vspace{-0.1cm}
We have performed experiments with three DG techniques: adversarial, multitasking, and MMD as discussed in Section~\ref{Sec:6}. All these approaches extend the state-of-the-art ECAPA-TDNN architecture and jointly optimize multiple cost functions with the trade-off parameter $\lambda$. Following~\citep{adi2019reverse}, we set $\lambda$ experimentally for better convergence of training. For the adversarial method, we have kept $\lambda = 0.001$ for the initial 15 epochs. For the other two methods, the initial value of $\lambda$ is $0.1$ for the first five epochs. After keeping the initial $\lambda$ value, we keep on increasing the $\lambda$ by $0.01$ per epoch for all the DG approaches. We have shown the comparative LID performances of different DG methods in Table.~\ref{tab:adversarial_LID_results}. \textcolor{black}{In this table, for comparative assessments, we have also reported the LID performance of augmented data generated by sampling utterances (with $\gamma=2$) from different signal-level augmentation categories.} %For assessing the relative improvement of the proposed MD-MMD method, we have also presented LID performance with the conventional RBF kernel MMD-loss.  

\begin{table}
\centering
\caption{LID performance (in EER~($\%$) / $C_{\mathrm{avg}} * 100$) of the proposed domain generalization approaches. \textcolor{black}{For comparison, LID performance with the sampled (with $\gamma = 2$) signal-level augmented utterances are also presented.}}
\label{tab:adversarial_LID_results}
\resizebox{\textwidth}{!}{%
\begin{tabular}{|c|rrr|lll|lll|} 
\hline
\multirow{3}{*}{Approaches} & \multicolumn{3}{c|}{Training corpus: IIITH} & \multicolumn{3}{c|}{Training corpus: LDC} & \multicolumn{3}{c|}{Training corpus: KGP} \\ 
\cline{2-10}
 & \multicolumn{1}{c}{IIITH-test} & \multicolumn{1}{c}{LDC-test} & \multicolumn{1}{c|}{KGP-test} & \multicolumn{1}{c}{IIITH-test} & \multicolumn{1}{c}{LDC-test} & \multicolumn{1}{c|}{KGP-test} & \multicolumn{1}{c}{IIITH-test} & \multicolumn{1}{c}{LDC-test} & \multicolumn{1}{c|}{KGP-test} \\ 
\cline{2-10}
 & \multicolumn{1}{c}{EER / $C_{\mathrm{avg}}$} & \multicolumn{1}{c}{EER / $C_{\mathrm{avg}}$} & \multicolumn{1}{c|}{EER / $C_{\mathrm{avg}}$} & \multicolumn{1}{c}{EER / $C_{\mathrm{avg}}$} & \multicolumn{1}{c}{EER / $C_{\mathrm{avg}}$} & \multicolumn{1}{c|}{EER / $C_{\mathrm{avg}}$} & \multicolumn{1}{c}{EER / $C_{\mathrm{avg}}$} & \multicolumn{1}{c}{EER / $C_{\mathrm{avg}}$} & \multicolumn{1}{c|}{EER / $C_{\mathrm{avg}}$} \\ 
\hline
Augmentation ($\gamma = 2$) & 8.00 / 9.14 & 38.19 / 42.71 & 38.88 / 35.52 & \textbf{32.91} / 32.53 & \textbf{20.10} / \textbf{24.22} & 48.25 / 41.53 & \textbf{33.67} / \textbf{33.79} & \textbf{47.25} / 45.49 & 8.94 / \textbf{7.26} \\
Adversarial & 9.64 / 10.88 & \textbf{35.40} / \textbf{40.68} & \textbf{28.28} / \textbf{22.92} & 34.98 / 36.77 & 20.80 / 25.14 & 44.94 / 43.59 & 36.76 / 34.35 & 48.26 / 42.73 & 9.24 / 8.98 \\
Multitasking & \textbf{7.97} / \textbf{9.03} & 36.29 / 41.89 & 29.47 / 25.69 & 36.32 / 36.40 & 21.28 / 25.47 & 44.27 / 44.50 & 38.97 / 36.12 & 48.12 / \textbf{42.59} & \textbf{8.79} / 8.70 \\
MMD & 9.06 / 9.78 & 39.58 / 45.30 & 33.15 / 26.28 & 38.89 / 35.81 & 22.10 / 26.02 & 46.32 / 45.88 & 40.58 / 38.15 & 48.44 / 45.05 & 11.98 / 13.70 \\
MD-MMD & 9.61 / 10.99 & 36.45 / 41.68 & 32.56 / 32.68 & 38.61 / \textbf{32.41} & 21.07 / 24.46 & \textbf{39.32} / \textbf{39.94} & 40.63 / 34.73 & 48.47 / 44.46 & 11.71 / 12.00 \\
\hline
\end{tabular}
}
\vspace{-.3cm}
\end{table}

We have observed a trend that the multitasking method yields the best performance for the same-corpora condition. For the cross-corpora evaluations, the adversarial model is found to be the most effective. The evaluation utterances of the same-corpora should have higher degrees of domain similarity with the training data. Therefore, the additional insights about the pseudo-domains learned from domain-aware language learning, improve the LID performance for such cases. Whereas the cross-corpora domains can be even more diverse, and in such cases incorporating domain invariance is more useful. For the IIITH-trained model, the performance improvements for KGP test data are more compared to the LDC test data. It can be justified as the meta information reveals the IIITH and KGP data are relatively similar in various aspects compared to the LDC data. The proposed MD-MMD shows prominent performance improvements as compared to the conventional MMD, especially for the cross-corpora scenarios. The MD-MMD method treats each pseudo-domain separately and weighs them as per the diversity distances. It enhances both same-corpora and cross-corpora LID performance. \textcolor{black}{MMD-based methods aim to reduce the distribution mismatch between the source (here, non-augmented original training data) and the target (here, augmented data with $\gamma=2$). It is reported in the existing literature that, while focusing on reducing the source-target distribution mismatch, MMD-based methods reduce the inter-class separability~\citep{yan2017mind,chen2019graph}. This can be the reason why both the two MMD-based methods do not outperform the adversarial or multitask learning based DG approaches in most cases.}

The explored DG approaches clearly outperform the baseline system for both same-corpora and cross-corpora evaluations. Considering the IIITH-trained system, from the baseline (in Table~\ref{tab:tdnn_eval}) $C_{\mathrm{avg}}$ improves by $5.55\%$ for the LDC-test utterances. The same for the KGP test utterances is $6.99\%$. Compared to the ECAPA-TDNN model, the corresponding relative improvements in $C_{\mathrm{avg}}$ are $4.14\%$ and $9.70\%$, respectively, for the LDC and KGP test corpora. The DG approaches further outperform the explored domain diversification approaches in most cases. The DG approaches utilize the augmented training data as pseudo-domains. So, we compare them with the ECAPA-TDNN model trained on augmented training set ($\gamma =2 $). The comparison reveals a relative $C_{\mathrm{avg}}$ improvement of $2.03\%$ and $12.60\%$, respectively, for the LDC and KGP test corpora. \textcolor{black}{If we study the LID performances for the different experiments we conduct, we can observe a clear trend that whenever LDC/KGP trained LID models are evaluated on KGP/LDC test utterances, the performance remains inferior compared to other cross-corpora train-test pairs. The meta information comparison, presented in Table~\ref{tab:corpora_comparison}, reveals a prominently higher corpora mismatch between these two corpora. For example, KGP is the smallest database, and its utterances are collected from broadcast news recordings. Whereas LDC is the largest database containing conversational telephone speech (CTS) data. The LDC utterances have a strong influence on regional dialects, and they are recorded in diverse backgrounds. Therefore, achieving cross-corpora generalization between such diverse corpora-pair is challenging. The language-specific cross-corpora study can provide further insights for solving this challenging situation. Still, compared to the baseline, we get the best-case EER improvements of $4.55\%$ for the LDC-test and $6.28\%$ for KGP-test data. On the other hand, for both same-corpora and cross-corpora scenarios, IIITH utterances show consistent performance improvements. IIITH contains both BN and CTS data, and it has a moderately large size. We believe the meta characteristics of this corpus have some commonalities with both corpora. So, achieving cross-corpora generalization over the IIITH corpus is relatively easier.} Overall, the LID performance improvements we achieve are promising, considering the fact that the corpora we use are low-resourced and developed independently with distinct differences in the meta-data characteristics.

\section{Further analysis and discussion}
\label{Sec:8}

In this section, we provide a blind cross-corpora evaluation of the proposed approaches on another unseen corpus, additional analysis, and ablation studies. We also discuss several prospects of the current study.

\begin{figure}[!ht]
%\vspace{-.4cm}
    \centering
    \includegraphics[trim=10cm 0.5cm 10cm 1.2cm,clip,width=.96\textwidth]{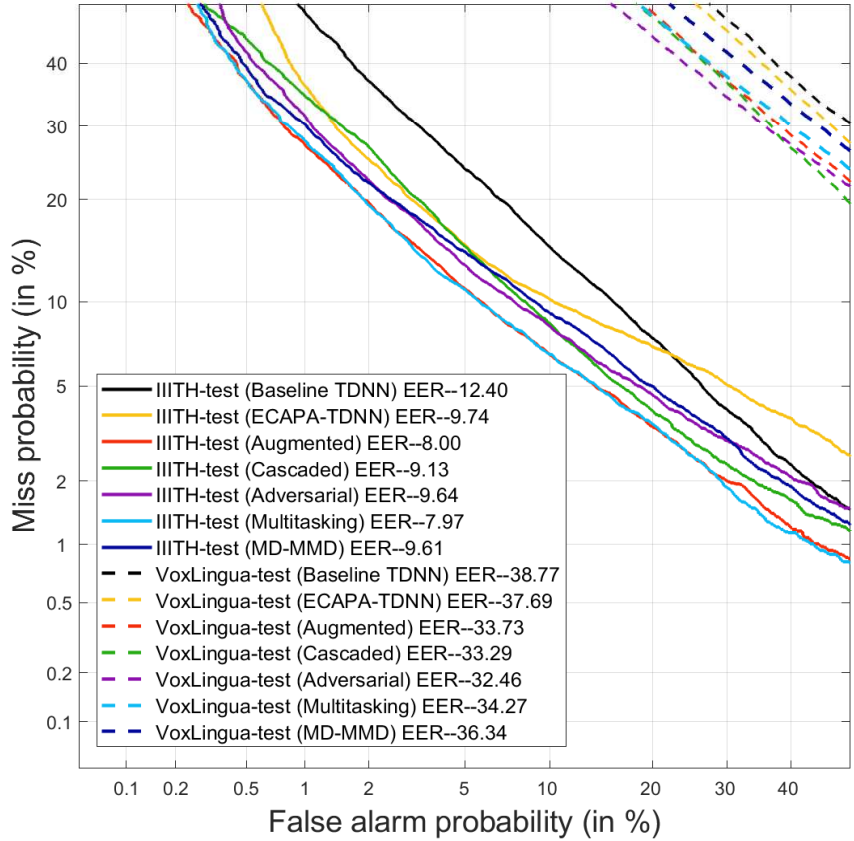}
    %\vspace{-.35cm}
    \caption{DET plots of the baselines and proposed approaches for the same-corpora (IIITH-test) and unseen cross-corpora (VoxLingua-test) conditions.}
    \label{fig:blind}
%    \vspace{-.38cm}
\end{figure}

\subsection{Evaluation on unseen corpus}
% \vspace{-0.1cm}
We have conducted extensive experiments so far on the three corpora and have demonstrated substantial improvements in cross-corpora experiments. This is indeed a valid argument that a machine learning algorithm should ideally refrain from using any kind of feedback from the evaluation set to optimize its configurations. Rather, the optimization should be made solely on the validation set. However, one may argue that the configuration and parameters of the proposed methods are optimized for the selected three corpora. Therefore, to further ensure the generalization ability of the proposed methods, we perform experiments on a fourth corpus \emph{VoxLingua107} (VoxLingua)~\citep{valk2021voxlingua107}, which is completely unseen during the system development process. VoxLingua is a vast and diverse corpus, with utterances automatically collected from wild sources using online video streaming platforms. We have only conducted the final evaluations with this corpus by randomly selecting 500 utterances of 3~s duration from each of the five languages. This likely includes speech utterances of diverse audio quality with various background noises that represent evaluation in the wild.

For this blind evaluation, we have considered the LID models trained on the IIITH corpus. Fig.~\ref{fig:blind} shows the \emph{detection error trade-off} (DET) curve as well as EERs for the same-corpora and cross-corpora evaluations following the different experiments, discussed earlier. We observe that all the explored approaches considerably outperform the baseline. Compared to the blind cross-corpora performance of the ECAPA-TDNN model, we achieve relative EER improvements of $4.40\%$ and $5.23\%$ for domain diversification and domain generalization methods, respectively. Among the domain diversification experiments, the proposed diversity-aware cascaded augmentation performs the best in the blind evaluation. Whereas the adversarial learning method outperforms all other DG methods for this evaluation.

\subsection{Assessing domain diversification}
We have proposed the maximally diversity-aware cascaded augmentations for effective diversification of the training data that can improve the cross-corpora generalization. We further analyze here whether the domain diversity is improved due to such augmentation methods. The domain diversity measures adopted in this work are data-driven, where we use embeddings from the LID model trained on augmented data. We now conduct this diversity analysis with a LID-independent model trained on different audio data. We select a pre-trained model trained for acoustic scene classification tasks. In particular, we choose \emph{OpenL3} embedding to explore the background and environmental differences across the three corpora~\citep {grollmisch2021analyzing}. 

\begin{figure}[!hbt]
%\vspace{-.15cm}
    \centering
    \includegraphics[trim=.1cm .2cm .2cm 0.2cm,clip,width=\linewidth]{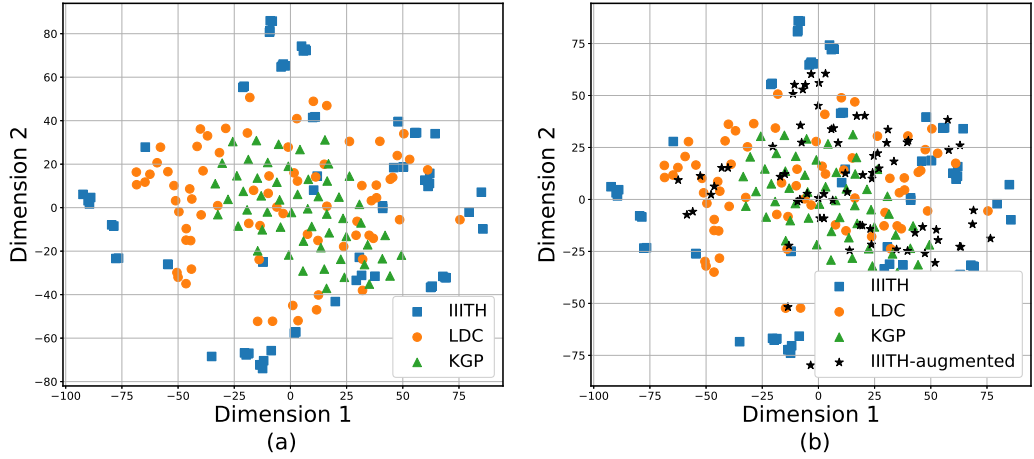}
    \vspace{-.15cm}
    \caption{T-SNE visualization of OpenL3 embeddings for (a) the training sets of the IIITH, LDC, and KGP corpora, (b) with further inclusion of the IIITH-augmented ($\gamma=2$) set.}
 %   \vspace{-.05cm}
    \label{fig:openl3}
  %  \vspace{-.25cm}
\end{figure}

Fig.~\ref{fig:openl3} visualizes the non-augmented training set embeddings for the three databases excluding and including IIITH augmented training set. We observe that three corpora consisting of non-augmented original training data (Fig.~\ref{fig:openl3} (a)) are somehow separated by the OpenL3 embeddings, and this indicates the presence of strong non-lingual differences.  We observe that the OpenL3 embeddings of the IIITH-augmented utterances (Fig.~\ref{fig:openl3} (b)) are well distributed across all three corpora demonstrating effective domain diversification with the proposed method. 

\begin{figure}[!hbt]
%\vspace{-.25cm}
    \centering
    \includegraphics[trim=.2cm 0cm .2cm 0cm,clip,width=\linewidth]{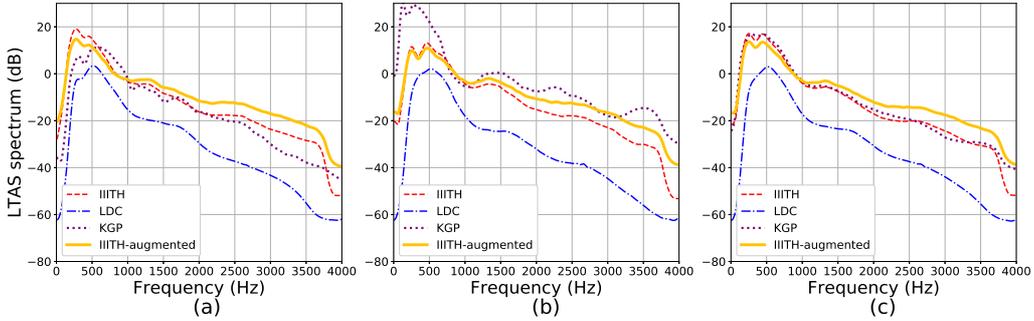}
    %\vspace{-.65cm}
    \caption{LTAS comparison of the three corpora and the IIITH-augmented data (with $\gamma=2$) for (a) Bengali, (b) Punjabi, and (c) all training utterances.}
    \label{fig:ltas_aug}
    %\vspace{-.25cm}
\end{figure}

We have also analyzed the overall LTAS of the three databases along with the augmented IIITH LTAS in Fig.~\ref{fig:ltas_aug}. The augmented spectrum shows closer proximity with the KGP-LTAS spectrum, indicating improved cross-corpora performance. We also present a language-specific LTAS comparison using two languages: Bengali and Punjabi. These two plots indicate that the cross-corpora mismatch is non-uniform across the languages. It promises a potential future extension of this work with language-specific cross-corpora performance analysis and generalization.

\subsection{\textcolor{black}{Linguistic diversity v/s non-lingual diversity in domain diversification}}
\label{sec:new}
\textcolor{black}{ In Section~\ref{Sec:6c}, we discuss the impact of the fold-factor, i.e. the ratio of the number of augmented utterances to the number of non-augmented original training utterances in the training set. In this case, we follow the common practice for augmented file selection as discussed in the literature~\citep{snyder2018spoken,garcia2020magneto}. This involves the generation of augmented copies for each method, followed by a random selection of augmented files from the pooled data to meet the desired fold-factor. Due to the random sampling from the pooled data, one utterance can have multiple augmented copies, i.e., content is repeated multiple times with varying non-lingual diversities. Whereas some of the utterances may not have any augmentation copy at all. In this section, we further explore the impact of content vs. non-lingual diversity within the augmented data. Such studies are not widely explored in the existing audio augmentation literature, especially for speaker and language recognition problems. Hence we find the inclusion of such discussion important for the augmentation research community.}

\textcolor{black}{We create three augmentation sets (with $\gamma=1$) using three scenarios. In the \emph{first scenario}, we follow the usual convention of sampling the augmented data, which is discussed at the beginning of this section and is followed in Section~\ref{sec:fold}. As mentioned above, in this scenario, there is a possibility of repetition of some content, whereas, for some utterances, there is a possibility of having no augmented copy. In the \emph{second scenario}, we apply a constraint that only one augmented version for each audio file is randomly selected from all augmented versions of the same file. Note that both scenarios consist of an equal number of audio files (twice the number of files in the original data due to  $\gamma=1$), but the former has lesser content variability than the latter. We also consider a \emph{third scenario} which further restricts the content variability by randomly selecting $1/2$ of the original utterances. Thereafter, for each of the selected utterances, we randomly select any three augmented copies (from seven different signal-level augmentation categories). The third augmentation set also contains an equal number of  total audio files as the other two conditions, but it has lesser content variability and more non-lingual variability.}

\begin{table}[!t]
\centering
\caption{\textcolor{black}{Impact of same-corpora and cross-corpora LID performance (in EER~($\%$) / $C_{\mathrm{avg}} * 100$)) for the content vs. non-lingual diversities in the augmented training sets. Here, three augmentation sets are created using the IIITH corpus, each having different degrees of content and non-lingual diversities.}}
\label{tab:content}
\resizebox{.7\linewidth}{!}{%
\begin{tabular}{|c|c|c|c|} 
\hline
\multirow{2}{*}{\textcolor{black}{Augmentation}} & \textcolor{black}{IIITH} & \textcolor{black}{LDC} & \textcolor{black}{KGP} \\ 
\cline{2-4}
 & \textcolor{black}{EER}/ \textcolor{black}{$C_\mathrm{avg}$} & \textcolor{black}{EER}/ \textcolor{black}{$C_\mathrm{avg}$} & \textcolor{black}{EER}/ \textcolor{black}{$C_\mathrm{avg}$} \\ 
\hline
\textcolor{black}{First scenario} & \textcolor{black}{8.75} / \textcolor{black}{10.01} & \textcolor{black}{\textbf{39.61}} / \textcolor{black}{43.96} & \textcolor{black}{38.65} / \textcolor{black}{36.70} \\ 
\hline
\textcolor{black}{Second scenario} & \textcolor{black}{7.40} / \textcolor{black}{9.41} & \textcolor{black}{42.31} / \textcolor{black}{45.99} & \textcolor{black}{\textbf{28.00}} / \textcolor{black}{\textbf{29.75}} \\ 
\hline
\textcolor{black}{Third scenario} & \textcolor{black}{\textbf{7.00}} / \textcolor{black}{\textbf{8.98}} & \textcolor{black}{46.15} / \textcolor{black}{\textbf{41.55}} & \textcolor{black}{31.00} / \textcolor{black}{30.37} \\
\hline
\end{tabular}
}
\end{table}

\textcolor{black}{Table~\ref{tab:content} shows the LID results for three augmentation scenarios with IIITH as the training corpus. Due to the randomness in audio file selection, we have conducted each experiment five times and we have reported the average performance. The results in Table~\ref{tab:content} show that applying constraints on the content repetition helps to improve the LID performance for the two test conditions: IIITH and KGP. The meta information says that IIITH and KGP both are mainly broadcast news data. So, the manner of articulation, accent, and dialect remains standard for these databases. Whereas LDC being conversational data, should contain diverse variations in accents and dialects. For such a database in evaluation, the constraint on content repetition during data augmentation does not help. We consider further exploration of content vs. non-lingual diversity as an interesting future work in the context of domain diversification.}

\subsection{Domain-invariant vs. domain-aware approaches}
%\vspace{-0.1cm}
The LID performance of the explored domain generalization methods reveals a trend that the adversarial models are suitable for cross-corpora generalization while the multitask learning models help to improve same-corpora LID performance. Therefore, we have two apparently opposite approaches to address the domain information during language training: \emph{domain-invariant} and \emph{domain-aware}~\citep{adi2019reverse}. The same-corpora data is more likely to have relatively higher domain similarity with the created pseudo-domains compared to the cross-corpora utterances. So, the additional task of pseudo-domain classification in the multitask learning architecture potentially makes the LID model aware of the possible diversities in the same-corpora test utterances. On the contrary, the domains of the cross-corpora evaluation utterances could be less similar to the pseudo-domains. In such scenarios, the domain-invariant approach helps improve the cross-corpora LID performance by making the model less sensitive to domain diversity.  Future work could consider exploring an algorithm for dynamically selecting either domain-aware or domain-invariant models for a given test audio~\citep{adi2019reverse}.

\subsection{Cross-corpora performance gap: Remaining challenges and solutions}
The proposed methods in this work show substantial improvements in cross-corpora evaluations. However, the performance gap between the same-corpora and cross-corpora evaluations is still considerably high. 
This could be due to several reasons: (i)~ The data augmentation methods used for domain diversification do not accurately simulate the real-world domains, and the audio data simulation is itself a challenging task~\citep{ribas2016study}. The domain generalization problem formulated in this study also relies on augmented data. We speculate that augmented data with more realistic conditions would help to reduce the performance gap. (ii)~Given the constraints in our problem formulation, we do not use any target domain data, and this limits the scope of optimization of the models for one specific domain. (iii)~The domain-invariant approach, which works well for the cross-corpora condition, does not entirely remove the domain-related characteristics from the latent representation. 
However, the main motivation of the current study is to assess the limits of domain diversification and domain generalization in challenging cross-corpora conditions with a focus on relative improvement over the existing baseline.  Following the short-duration NIST LRE evaluation conditions~\citep{sadjadi20182017}, we also use a very short test speech of duration 3~s, which makes the LID problem more difficult~\citep{gonzalez2015frame}. The state-of-the-art LID systems show about $15\%$ EER on 3~s duration condition of LRE17 where the systems are both trained and tested on conversational telephone speech~\citep{padi2020towards}. Even though the systems with such higher EERs are unreliable, we believe that the relative improvements we achieve provide a promising direction for solving this challenging task of cross-corpora generalization with short-duration test speech.

In the current work, the application of the proposed domain diversification and domain generalization is limited to the ECAPA-TDNN, a specific type of model for language classification which trains architecture in a discriminative way. The study can be extended with other discriminative and generative approaches. Recently, self-supervised learning (SSL)~\citep{mohamed2022self} are being widely explored for different tasks, including language recognition~\citep{liu2022efficient}, which exploits a large amount of unlabeled audio data from other sources. We consider cross-corpora evaluation with SSL frameworks and assessing the performance gaps with same-corpora evaluation as interesting future research avenues.

\section{Conclusions}
\label{Sec:9}
 This study has explored the suitability of domain diversification and domain generalization for improving cross-corpora generalization in low-resourced LID tasks. We have explored existing and new data audio data augmentation methods for domain diversification. For domain generalization, we investigate adversarial learning, multitask learning as well as MMD loss, and its new variants with pseudo-domains. In addition to extensive experiments on three widely used Indian LID corpora, we also perform a blind evaluation on an unseen corpus to further validate the effectiveness of the proposed approaches. We summarize the main findings of this work as follows: 

\begin{itemize}
    \item Different evaluation corpora can contain diverse non-lingual characteristics. Hence, along with the conventional audio augmentations, exploring new sources of perturbing the audio data further diversify the training data, which is effective for unseen cross-corpora generalization. For example, the proposed speech enhancement or codec-based augmentations are most promising for improving robustness against LDC and KGP test data for the IIITH-trained LID models. 

    \item The fundamental motivation for applying domain diversification is to reduce the cross-corpora mismatch by simulating the non-lingual characteristics related to the potential unseen domain. Hence, augmenting audio data in a more realistic manner is important. For this reason, we propose the maximally diversity-aware cascaded augmentation with an optimized fold-factor, which is a more effective solution in blind evaluation than the conventional augmentation methods.
    
    \item We aim to achieve better LID generalization across diverse unseen domains with limited training data. Hence, instead of attaining robustness for a specific target domain using the conventional domain adaptation methods, we opt for domain generalization. It utilizes the different augmentation categories as pseudo-domains to further improve the LID performance compared to the domain diversification methods.
    
    \item We explore domain generalization in two fundamentally opposite ways using the domain-invariant and domain-aware methods. Domain-invariant LID models are trained using gradient reversal or MMD-based methods in such a way that it becomes less sensitive to non-lingual domain-specific mismatches. In the domain-aware cases, we rather update the LID model to additionally learn domain-specific cues using multitask learning. Our study reveals that domain-invariant learning is more effective for cross-corpora generalization, whereas domain-awareness improves same-corpora LID performance.
\end{itemize}

Although the proposed domain diversification and domain generalization largely improve the cross-corpora LID performance, investigations are still required to reduce the same-corpora and cross-corpora performance gap further. Exploring other advanced data augmentation methods using voice conversion, text-to-speech synthesis, and style transfer could be promising. We address the cross-corpora problem from the perspectives of the entire database. Investigating the cross-corpora LID in a language-specific manner can reveal further insights into this problem. \textcolor{black}{Our further investigation of the spoken content v/s non-lingual variability indicates another promising future direction.} 

\section*{Acknowledgements}
\textcolor{black}{The first author would also like to thank the Linguistic Data Consortium (LDC), University of Pennsylvania, USA, for the LDC Data Scholarship. We also thank Prof. KS Rao (IIT Kharagpur) and Prof. AK Vuppala (IIIT Hyderabad) for their help with the KGP and IIITH datasets. Finally, the authors would like to thank the associate editor and the reviewers for their careful reading, detailed comments, and constructive suggestions, which substantially enhanced the content of the manuscript.}

\bibliographystyle{cas-model2-names}
\bibliography{main}

\begin{thebibliography}{106}
\expandafter\ifx\csname natexlab\endcsname\relax\def\natexlab#1{#1}\fi
\providecommand{\url}[1]{\texttt{#1}}
\providecommand{\href}[2]{#2}
\providecommand{\path}[1]{#1}
\providecommand{\DOIprefix}{doi:}
\providecommand{\ArXivprefix}{arXiv:}
\providecommand{\URLprefix}{URL: }
\providecommand{\Pubmedprefix}{pmid:}
\providecommand{\doi}[1]{\href{http://dx.doi.org/#1}{\path{#1}}}
\providecommand{\Pubmed}[1]{\href{pmid:#1}{\path{#1}}}
\providecommand{\bibinfo}[2]{#2}
\ifx\xfnm\relax \def\xfnm[#1]{\unskip,\space#1}\fi
%Type = Inproceedings
\bibitem[{Adi et~al.(2019)Adi, Zeghidour, Collobert, Usunier, Liptchinsky and
  Synnaeve}]{adi2019reverse}
\bibinfo{author}{Adi, Y.}, \bibinfo{author}{Zeghidour, N.},
  \bibinfo{author}{Collobert, R.}, \bibinfo{author}{Usunier, N.},
  \bibinfo{author}{Liptchinsky, V.}, \bibinfo{author}{Synnaeve, G.},
  \bibinfo{year}{2019}.
\newblock \bibinfo{title}{To reverse the gradient or not: an empirical
  comparison of adversarial and multi-task learning in speech recognition}, in:
  \bibinfo{booktitle}{ICASSP}, \bibinfo{organization}{IEEE}. pp.
  \bibinfo{pages}{3742--3746}.
%Type = Inproceedings
\bibitem[{Alumäe and Kukk(2022)}]{alumae22_odyssey}
\bibinfo{author}{Alumäe, T.}, \bibinfo{author}{Kukk, K.},
  \bibinfo{year}{2022}.
\newblock \bibinfo{title}{Pretraining approaches for spoken language
  recognition: Tal{T}ech submission to the {OLR} 2021 challenge}, in:
  \bibinfo{booktitle}{Odyssey: The Speaker and Language Recognition Workshop},
  \bibinfo{organization}{ISCA}. pp. \bibinfo{pages}{240--247}.
%Type = Article
\bibitem[{Benyassine et~al.(1997)Benyassine, Shlomot, Su, Massaloux, Lamblin
  and Petit}]{benyassine1997silence}
\bibinfo{author}{Benyassine, A.}, \bibinfo{author}{Shlomot, E.},
  \bibinfo{author}{Su, H.}, \bibinfo{author}{Massaloux, D.},
  \bibinfo{author}{Lamblin, C.}, \bibinfo{author}{Petit, J.},
  \bibinfo{year}{1997}.
\newblock \bibinfo{title}{A silence compression scheme for use with g. 729
  optimized for v. 70 digital simultaneous voice and data applications
  (recommendation g. 729 annex b)}.
\newblock \bibinfo{journal}{IEEE {C}ommun. {M}ag} \bibinfo{volume}{35},
  \bibinfo{pages}{64--73}.
%Type = Inproceedings
\bibitem[{Berouti et~al.(1979)Berouti, Schwartz and
  Makhoul}]{berouti1979enhancement}
\bibinfo{author}{Berouti, M.}, \bibinfo{author}{Schwartz, R.},
  \bibinfo{author}{Makhoul, J.}, \bibinfo{year}{1979}.
\newblock \bibinfo{title}{Enhancement of speech corrupted by acoustic noise},
  in: \bibinfo{booktitle}{ICASSP}, \bibinfo{organization}{IEEE}. pp.
  \bibinfo{pages}{208--211}.
%Type = Article
\bibitem[{Beyan et~al.(2021)Beyan, Shahid and Murino}]{9133504}
\bibinfo{author}{Beyan, C.}, \bibinfo{author}{Shahid, M.},
  \bibinfo{author}{Murino, V.}, \bibinfo{year}{2021}.
\newblock \bibinfo{title}{Real{VAD}: A real-world dataset and a method for
  voice activity detection by body motion analysis}.
\newblock \bibinfo{journal}{IEEE Transactions on Multimedia}
  \bibinfo{volume}{23}, \bibinfo{pages}{2071--2085}.
%Type = Article
\bibitem[{Blanchard et~al.(2011)Blanchard, Lee and
  Scott}]{blanchard2011generalizing}
\bibinfo{author}{Blanchard, G.}, \bibinfo{author}{Lee, G.},
  \bibinfo{author}{Scott, C.}, \bibinfo{year}{2011}.
\newblock \bibinfo{title}{Generalizing from several related classification
  tasks to a new unlabeled sample}.
\newblock \bibinfo{journal}{Advances in {N}eural {I}nformation {P}rocessing
  {S}ystems} \bibinfo{volume}{24}, \bibinfo{pages}{2178--2186}.
%Type = Article
\bibitem[{Brookes et~al.(1997)}]{brookes1997voicebox}
\bibinfo{author}{Brookes, M.}, et~al., \bibinfo{year}{1997}.
\newblock \bibinfo{title}{Voicebox: Speech processing toolbox for {MATLAB}}.
\newblock \bibinfo{journal}{Software, available [Mar. 2011] from www. ee. ic.
  ac. uk/hp/staff/dmb/voicebox/voicebox.html} \bibinfo{volume}{47}.
%Type = Article
\bibitem[{Br{\"u}mmer and Du~Preez(2006)}]{brummer2006application}
\bibinfo{author}{Br{\"u}mmer, N.}, \bibinfo{author}{Du~Preez, J.},
  \bibinfo{year}{2006}.
\newblock \bibinfo{title}{Application-independent evaluation of speaker
  detection}.
\newblock \bibinfo{journal}{Computer Speech \& Language} \bibinfo{volume}{20},
  \bibinfo{pages}{230--275}.
%Type = Article
\bibitem[{Caruana(1997)}]{caruana1997multitask}
\bibinfo{author}{Caruana, R.}, \bibinfo{year}{1997}.
\newblock \bibinfo{title}{Multitask learning}.
\newblock \bibinfo{journal}{Machine learning} \bibinfo{volume}{28},
  \bibinfo{pages}{41--75}.
%Type = Inproceedings
\bibitem[{Cha et~al.(2021)Cha, Chun, Lee, Cho, Lee and Park}]{cha2021domain}
\bibinfo{author}{Cha, J.}, \bibinfo{author}{Chun, S.}, \bibinfo{author}{Lee,
  K.}, \bibinfo{author}{Cho, Han-Cheol~andPark, S.}, \bibinfo{author}{Lee, Y.},
  \bibinfo{author}{Park, S.}, \bibinfo{year}{2021}.
\newblock \bibinfo{title}{{SWAD}: Domain generalization by seeking flat
  minima}, in: \bibinfo{booktitle}{Advances in {N}eural {I}nformation
  {P}rocessing {S}ystems}, pp. \bibinfo{pages}{22405--22418}.
%Type = Inproceedings
\bibitem[{Chakraborty et~al.(2021)Chakraborty, Chakraborty and
  Bhattacharya}]{chakraborty2021denserecognition}
\bibinfo{author}{Chakraborty, J.}, \bibinfo{author}{Chakraborty, B.},
  \bibinfo{author}{Bhattacharya, U.}, \bibinfo{year}{2021}.
\newblock \bibinfo{title}{Dense{R}ecognition of spoken languages}, in:
  \bibinfo{booktitle}{International Conference on Pattern Recognition (ICPR)},
  \bibinfo{organization}{IEEE}. pp. \bibinfo{pages}{9674--9681}.
%Type = Article
\bibitem[{Chen et~al.(2019)Chen, Song, Li and Wu}]{chen2019graph}
\bibinfo{author}{Chen, Y.}, \bibinfo{author}{Song, S.}, \bibinfo{author}{Li,
  S.}, \bibinfo{author}{Wu, C.}, \bibinfo{year}{2019}.
\newblock \bibinfo{title}{A graph embedding framework for maximum mean
  discrepancy-based domain adaptation algorithms}.
\newblock \bibinfo{journal}{{IEEE} {T}ransactions on {I}mage {P}rocessing}
  \bibinfo{volume}{29}, \bibinfo{pages}{199--213}.
%Type = Inproceedings
\bibitem[{Chettri et~al.(2021)Chettri, Hautamäki, Sahidullah and
  Kinnunen}]{chettri2021data}
\bibinfo{author}{Chettri, B.}, \bibinfo{author}{Hautamäki, R.G.},
  \bibinfo{author}{Sahidullah, M.}, \bibinfo{author}{Kinnunen, T.},
  \bibinfo{year}{2021}.
\newblock \bibinfo{title}{{Data quality as predictor of voice anti-spoofing
  generalization}}, in: \bibinfo{booktitle}{INTERSPEECH},
  \bibinfo{organization}{ISCA}. pp. \bibinfo{pages}{1659--1663}.
%Type = Article
\bibitem[{Clark et~al.(2019)Clark, Doyle, Garaialde, Gilmartin, Schl{\"o}gl,
  Edlund, Aylett, Cabral, Munteanu, Edwards et~al.}]{clark2019state}
\bibinfo{author}{Clark, L.}, \bibinfo{author}{Doyle, P.},
  \bibinfo{author}{Garaialde, D.}, \bibinfo{author}{Gilmartin, E.},
  \bibinfo{author}{Schl{\"o}gl, S.}, \bibinfo{author}{Edlund, J.},
  \bibinfo{author}{Aylett, M.}, \bibinfo{author}{Cabral, J.},
  \bibinfo{author}{Munteanu, C.}, \bibinfo{author}{Edwards, J.}, et~al.,
  \bibinfo{year}{2019}.
\newblock \bibinfo{title}{The state of speech in {HCI}: Trends, themes and
  challenges}.
\newblock \bibinfo{journal}{Interacting with {C}omputers} \bibinfo{volume}{31},
  \bibinfo{pages}{349--371}.
%Type = Inproceedings
\bibitem[{Deng et~al.(2018)Deng, Zheng, Ye, Kang, Yang and
  Jiao}]{deng2018image}
\bibinfo{author}{Deng, W.}, \bibinfo{author}{Zheng, L.}, \bibinfo{author}{Ye,
  Q.}, \bibinfo{author}{Kang, G.}, \bibinfo{author}{Yang, Y.},
  \bibinfo{author}{Jiao, J.}, \bibinfo{year}{2018}.
\newblock \bibinfo{title}{Image-image domain adaptation with preserved
  self-similarity and domain-dissimilarity for person re-identification}, in:
  \bibinfo{booktitle}{CVPR}, pp. \bibinfo{pages}{994--1003}.
%Type = Inproceedings
\bibitem[{Desplanques et~al.(2020)Desplanques, Thienpondt and
  Demuynck}]{desplanques2020ecapa}
\bibinfo{author}{Desplanques, B.}, \bibinfo{author}{Thienpondt, J.},
  \bibinfo{author}{Demuynck, K.}, \bibinfo{year}{2020}.
\newblock \bibinfo{title}{{ECAPA}-{TDNN}: Emphasized channel attention,
  propagation and aggregation in {TDNN} based speaker verification}, in:
  \bibinfo{booktitle}{INTERSPEECH}, \bibinfo{organization}{ISCA}. pp.
  \bibinfo{pages}{1--5}.
%Type = Inproceedings
\bibitem[{Dey et~al.(2021)Dey, Saha and Sahidullah}]{dey2021cross}
\bibinfo{author}{Dey, S.}, \bibinfo{author}{Saha, G.},
  \bibinfo{author}{Sahidullah, M.}, \bibinfo{year}{2021}.
\newblock \bibinfo{title}{Cross-corpora language recognition: A preliminary
  investigation with {I}ndian languages}, in: \bibinfo{booktitle}{EUSIPCO},
  \bibinfo{organization}{IEEE}. pp. \bibinfo{pages}{546--550}.
%Type = Article
\bibitem[{Dey et~al.(2022)Dey, Sahidullah and Saha}]{dey2021overview}
\bibinfo{author}{Dey, S.}, \bibinfo{author}{Sahidullah, M.},
  \bibinfo{author}{Saha, G.}, \bibinfo{year}{2022}.
\newblock \bibinfo{title}{An overview on {I}ndian spoken language recognition
  from machine learning perspective.}
\newblock \bibinfo{journal}{ACM {T}ransactions on {A}sian and {L}ow-{R}esource
  {L}anguage {I}nformation {P}rocessing} \bibinfo{volume}{21},
  \bibinfo{pages}{1--45}.
%Type = Article
\bibitem[{Ding and Fu(2017)}]{ding2017deep}
\bibinfo{author}{Ding, Z.}, \bibinfo{author}{Fu, Y.}, \bibinfo{year}{2017}.
\newblock \bibinfo{title}{Deep domain generalization with structured low-rank
  constraint}.
\newblock \bibinfo{journal}{IEEE {T}ransactions on {I}mage {P}rocessing}
  \bibinfo{volume}{27}, \bibinfo{pages}{304--313}.
%Type = Article
\bibitem[{Doire et~al.(2016)}]{doire2016single}
\bibinfo{author}{Doire, C.S.}, et~al., \bibinfo{year}{2016}.
\newblock \bibinfo{title}{Single-channel online enhancement of speech corrupted
  by reverberation and noise}.
\newblock \bibinfo{journal}{IEEE/ACM {T}ransactions on {A}udio, {S}peech, and
  {L}anguage {P}rocessing} \bibinfo{volume}{25}, \bibinfo{pages}{572--587}.
%Type = Inproceedings
\bibitem[{Du et~al.(2021)Du, Li, Lu, Wang and Qian}]{du2021data}
\bibinfo{author}{Du, C.}, \bibinfo{author}{Li, H.}, \bibinfo{author}{Lu, Y.},
  \bibinfo{author}{Wang, L.}, \bibinfo{author}{Qian, Y.}, \bibinfo{year}{2021}.
\newblock \bibinfo{title}{Data augmentation for end-to-end code-switching
  speech recognition}, in: \bibinfo{booktitle}{Spoken Language Technology
  Workshop (SLT)}, \bibinfo{organization}{IEEE}. pp. \bibinfo{pages}{194--200}.
%Type = Inproceedings
\bibitem[{Duroselle et~al.(2020)Duroselle, Jouvet and
  Illina}]{duroselle20_interspeech}
\bibinfo{author}{Duroselle, R.}, \bibinfo{author}{Jouvet, D.},
  \bibinfo{author}{Illina, I.}, \bibinfo{year}{2020}.
\newblock \bibinfo{title}{Metric learning loss functions to reduce domain
  mismatch in the x-vector space for language recognition}, in:
  \bibinfo{booktitle}{INTERSPEECH}, pp. \bibinfo{pages}{447--451}.
%Type = Article
\bibitem[{Ferrer et~al.(2022)Ferrer, Castan, McLaren and
  Lawson}]{ferrer2022discriminative}
\bibinfo{author}{Ferrer, L.}, \bibinfo{author}{Castan, D.},
  \bibinfo{author}{McLaren, M.}, \bibinfo{author}{Lawson, A.},
  \bibinfo{year}{2022}.
\newblock \bibinfo{title}{A discriminative hierarchical {PLDA}-based model for
  spoken language recognition}.
\newblock \bibinfo{journal}{IEEE/ACM Transactions on Audio, Speech, and
  Language Processing} \bibinfo{volume}{30}, \bibinfo{pages}{2396--2410}.
%Type = Article
\bibitem[{Ganin et~al.(2016)Ganin, Ustinova, Ajakan, Germain, Larochelle,
  Laviolette, Marchand and Lempitsky}]{ganin2016domain}
\bibinfo{author}{Ganin, Y.}, \bibinfo{author}{Ustinova, E.},
  \bibinfo{author}{Ajakan, H.}, \bibinfo{author}{Germain, P.},
  \bibinfo{author}{Larochelle, H.}, \bibinfo{author}{Laviolette, F.},
  \bibinfo{author}{Marchand, M.}, \bibinfo{author}{Lempitsky, V.},
  \bibinfo{year}{2016}.
\newblock \bibinfo{title}{Domain-adversarial training of neural networks}.
\newblock \bibinfo{journal}{The {J}ournal of {M}achine {L}earning {R}esearch}
  \bibinfo{volume}{17}, \bibinfo{pages}{2096--2030}.
%Type = Inproceedings
\bibitem[{Garcia-Romero et~al.(2020)Garcia-Romero, Sell and
  Mccree}]{garcia2020magneto}
\bibinfo{author}{Garcia-Romero, D.}, \bibinfo{author}{Sell, G.},
  \bibinfo{author}{Mccree, A.}, \bibinfo{year}{2020}.
\newblock \bibinfo{title}{Mag{N}et{O}: X-vector magnitude estimation network
  plus offset for improved speaker recognition}, in:
  \bibinfo{booktitle}{Odyssey: The Speaker and Language Recognition Workshop},
  \bibinfo{organization}{ISCA}. pp. \bibinfo{pages}{1--8}.
%Type = Article
\bibitem[{Gerczuk et~al.(2021)Gerczuk, Amiriparian, Ottl and Schuller}]{EmoNet}
\bibinfo{author}{Gerczuk, M.}, \bibinfo{author}{Amiriparian, S.},
  \bibinfo{author}{Ottl, S.}, \bibinfo{author}{Schuller, B.},
  \bibinfo{year}{2021}.
\newblock \bibinfo{title}{{EmoNet}: A transfer learning framework for
  multi-corpus speech emotion recognition}.
\newblock \bibinfo{journal}{IEEE Transactions on Affective Computing} ,
  \bibinfo{pages}{1--1}.
%Type = Article
\bibitem[{Gerkmann and Hendriks(2011)}]{gerkmann2011unbiased}
\bibinfo{author}{Gerkmann, T.}, \bibinfo{author}{Hendriks, R.C.},
  \bibinfo{year}{2011}.
\newblock \bibinfo{title}{Unbiased {MMSE}-based noise power estimation with low
  complexity and low tracking delay}.
\newblock \bibinfo{journal}{IEEE {T}ransactions on {A}udio, {S}peech, and
  {L}anguage {P}rocessing} \bibinfo{volume}{20}, \bibinfo{pages}{1383--1393}.
%Type = Article
\bibitem[{Gideon et~al.(2021)Gideon, McInnis and Provost}]{gideon2021improving}
\bibinfo{author}{Gideon, J.}, \bibinfo{author}{McInnis, M.G.},
  \bibinfo{author}{Provost, E.M.}, \bibinfo{year}{2021}.
\newblock \bibinfo{title}{Improving cross-corpus speech emotion recognition
  with adversarial discriminative domain generalization ({ADD}o{G})}.
\newblock \bibinfo{journal}{IEEE {T}ransactions on {A}ffective {C}omputing}
  \bibinfo{volume}{12}, \bibinfo{pages}{1055--1068}.
%Type = Inproceedings
\bibitem[{Gillespie et~al.(2017)Gillespie, Logan, Moore, Laures-Gore, Russell
  and Patel}]{gillespie2017cross}
\bibinfo{author}{Gillespie, S.}, \bibinfo{author}{Logan, Y.Y.},
  \bibinfo{author}{Moore, E.}, \bibinfo{author}{Laures-Gore, J.},
  \bibinfo{author}{Russell, S.}, \bibinfo{author}{Patel, R.},
  \bibinfo{year}{2017}.
\newblock \bibinfo{title}{Cross-database models for the classification of
  dysarthria presence.}, in: \bibinfo{booktitle}{INTERSPEECH},
  \bibinfo{organization}{ISCA}. pp. \bibinfo{pages}{3127--3131}.
%Type = Article
\bibitem[{Gonzalez-Dominguez et~al.(2015)}]{gonzalez2015frame}
\bibinfo{author}{Gonzalez-Dominguez, J.}, et~al., \bibinfo{year}{2015}.
\newblock \bibinfo{title}{Frame-by-frame language identification in short
  utterances using deep neural networks}.
\newblock \bibinfo{journal}{Neural {N}etworks} \bibinfo{volume}{64},
  \bibinfo{pages}{49--58}.
%Type = Inproceedings
\bibitem[{Greenberg et~al.(2012)Greenberg, Martin and
  Przybocki}]{greenberg20122011}
\bibinfo{author}{Greenberg, C.S.}, \bibinfo{author}{Martin, A.F.},
  \bibinfo{author}{Przybocki, M.A.}, \bibinfo{year}{2012}.
\newblock \bibinfo{title}{The 2011 {NIST} language recognition evaluation}, in:
  \bibinfo{booktitle}{INTERSPEECH}, \bibinfo{organization}{ISCA}. pp.
  \bibinfo{pages}{34--37}.
%Type = Article
\bibitem[{Gretton et~al.(2006)Gretton, Borgwardt, Rasch, Sch{\"o}lkopf and
  Smola}]{gretton2006kernel}
\bibinfo{author}{Gretton, A.}, \bibinfo{author}{Borgwardt, K.},
  \bibinfo{author}{Rasch, M.}, \bibinfo{author}{Sch{\"o}lkopf, B.},
  \bibinfo{author}{Smola, A.}, \bibinfo{year}{2006}.
\newblock \bibinfo{title}{A kernel method for the two-sample-problem}.
\newblock \bibinfo{journal}{Advances in {N}eural {I}nformation {P}rocessing
  {S}ystems} \bibinfo{volume}{19}.
%Type = Inproceedings
\bibitem[{Grollmisch et~al.(2021)Grollmisch, Cano, Kehling and
  Taenzer}]{grollmisch2021analyzing}
\bibinfo{author}{Grollmisch, S.}, \bibinfo{author}{Cano, E.},
  \bibinfo{author}{Kehling, C.}, \bibinfo{author}{Taenzer, M.},
  \bibinfo{year}{2021}.
\newblock \bibinfo{title}{Analyzing the potential of pre-trained embeddings for
  audio classification tasks}, in: \bibinfo{booktitle}{EUSIPCO},
  \bibinfo{organization}{IEEE}. pp. \bibinfo{pages}{790--794}.
%Type = Inproceedings
\bibitem[{Gulrajani and Lopez-Paz(2021)}]{gulrajani2020search}
\bibinfo{author}{Gulrajani, I.}, \bibinfo{author}{Lopez-Paz, D.},
  \bibinfo{year}{2021}.
\newblock \bibinfo{title}{In search of lost domain generalization}, in:
  \bibinfo{booktitle}{International Conference on Learning Representations}.
%Type = Article
\bibitem[{Hu and Yang(2017)}]{7395312}
\bibinfo{author}{Hu, X.}, \bibinfo{author}{Yang, Y.H.}, \bibinfo{year}{2017}.
\newblock \bibinfo{title}{Cross-dataset and cross-cultural music mood
  prediction: A case on {W}estern and {C}hinese pop songs}.
\newblock \bibinfo{journal}{IEEE {T}ransactions on {A}ffective {C}omputing}
  \bibinfo{volume}{8}, \bibinfo{pages}{228--240}.
%Type = Inproceedings
\bibitem[{Iqbal et~al.(2021)Iqbal, Helwani, Krishnaswamy and
  Wang}]{iqbal2021enhancing}
\bibinfo{author}{Iqbal, T.}, \bibinfo{author}{Helwani, K.},
  \bibinfo{author}{Krishnaswamy, A.}, \bibinfo{author}{Wang, W.},
  \bibinfo{year}{2021}.
\newblock \bibinfo{title}{Enhancing audio augmentation methods with consistency
  learning}, in: \bibinfo{booktitle}{ICASSP}, \bibinfo{organization}{IEEE}. pp.
  \bibinfo{pages}{646--650}.
%Type = Inproceedings
\bibitem[{Kang et~al.(2022)Kang, Alam and Fathan}]{kang2022deep}
\bibinfo{author}{Kang, W.}, \bibinfo{author}{Alam, M.J.},
  \bibinfo{author}{Fathan, A.}, \bibinfo{year}{2022}.
\newblock \bibinfo{title}{Deep learning-based end-to-end spoken language
  identification system for domain-mismatched scenario}, in:
  \bibinfo{booktitle}{Language {R}esources and {E}valuation {C}onference}, pp.
  \bibinfo{pages}{7339--7343}.
%Type = Article
\bibitem[{Karen et~al.(2017)}]{LDC}
\bibinfo{author}{Karen, J.}, et~al., \bibinfo{year}{2017}.
\newblock \bibinfo{title}{Multi-language conversational telephone speech 2011
  -- {S}outh {A}sian {LDC2017S14}. web download. {P}hiladelphia: {L}inguistic
  {D}ata {C}onsortium}.
\newblock \bibinfo{journal}{-} .
%Type = Inproceedings
\bibitem[{Khosla et~al.(2012)Khosla, Zhou, Malisiewicz, Efros and
  Torralba}]{khosla2012undoing}
\bibinfo{author}{Khosla, A.}, \bibinfo{author}{Zhou, T.},
  \bibinfo{author}{Malisiewicz, T.}, \bibinfo{author}{Efros, A.A.},
  \bibinfo{author}{Torralba, A.}, \bibinfo{year}{2012}.
\newblock \bibinfo{title}{Undoing the damage of dataset bias}, in:
  \bibinfo{booktitle}{European Conference on Computer Vision},
  \bibinfo{organization}{Springer}. pp. \bibinfo{pages}{158--171}.
%Type = Incollection
\bibitem[{Korshunov and Marcel(2019)}]{korshunov2019cross}
\bibinfo{author}{Korshunov, P.}, \bibinfo{author}{Marcel, S.},
  \bibinfo{year}{2019}.
\newblock \bibinfo{title}{A cross-database study of voice presentation attack
  detection}, in: \bibinfo{booktitle}{Handbook of Biometric Anti-Spoofing}.
  \bibinfo{publisher}{Springer}, pp. \bibinfo{pages}{363--389}.
%Type = Inproceedings
\bibitem[{Kumawat and Routray(2021)}]{kumawat2021applying}
\bibinfo{author}{Kumawat, P.}, \bibinfo{author}{Routray, A.},
  \bibinfo{year}{2021}.
\newblock \bibinfo{title}{Applying {TDNN} architectures for analyzing duration
  dependencies on speech emotion recognition}, in:
  \bibinfo{booktitle}{INTERSPEECH}, \bibinfo{organization}{ISCA}. pp.
  \bibinfo{pages}{3410--3414}.
%Type = Article
\bibitem[{Li et~al.(2013)Li, Ma and Lee}]{li2013spoken}
\bibinfo{author}{Li, H.}, \bibinfo{author}{Ma, B.}, \bibinfo{author}{Lee,
  K.A.}, \bibinfo{year}{2013}.
\newblock \bibinfo{title}{Spoken language recognition: from fundamentals to
  practice}.
\newblock \bibinfo{journal}{Proceedings of the IEEE} \bibinfo{volume}{101},
  \bibinfo{pages}{1136--1159}.
%Type = Article
\bibitem[{Li et~al.(2021)Li, Li, Liu and Hong}]{li2021deep}
\bibinfo{author}{Li, L.}, \bibinfo{author}{Li, Z.}, \bibinfo{author}{Liu, Y.},
  \bibinfo{author}{Hong, Q.}, \bibinfo{year}{2021}.
\newblock \bibinfo{title}{Deep joint learning for language recognition}.
\newblock \bibinfo{journal}{Neural {N}etworks} \bibinfo{volume}{141},
  \bibinfo{pages}{72--86}.
%Type = Inproceedings
\bibitem[{Li et~al.(2020)Li, Zhao, Hong, Li, Tang, Wang, Song and
  Yang}]{li2020ap20}
\bibinfo{author}{Li, Z.}, \bibinfo{author}{Zhao, M.}, \bibinfo{author}{Hong,
  Q.}, \bibinfo{author}{Li, L.}, \bibinfo{author}{Tang, Z.},
  \bibinfo{author}{Wang, D.}, \bibinfo{author}{Song, L.},
  \bibinfo{author}{Yang, C.}, \bibinfo{year}{2020}.
\newblock \bibinfo{title}{{AP20-OLR} challenge: Three tasks and their
  baselines}, in: \bibinfo{booktitle}{Asia-Pacific Signal and Information
  Processing Association Annual Summit and Conference (APSIPA ASC)},
  \bibinfo{organization}{IEEE}. pp. \bibinfo{pages}{550--555}.
%Type = Inproceedings
\bibitem[{Liu et~al.(2022a)Liu, {Garcia Perera}, Khong, Styles and
  Khudanpur}]{liu22e_interspeech}
\bibinfo{author}{Liu, H.}, \bibinfo{author}{{Garcia Perera}, L.P.},
  \bibinfo{author}{Khong, A.}, \bibinfo{author}{Styles, S.},
  \bibinfo{author}{Khudanpur, S.}, \bibinfo{year}{2022}a.
\newblock \bibinfo{title}{{PHO-LID}: A unified model incorporating
  acoustic-phonetic and phonotactic information for language identification},
  in: \bibinfo{booktitle}{INTERSPEECH}, \bibinfo{organization}{ISCA}. pp.
  \bibinfo{pages}{2233--2237}.
%Type = Article
\bibitem[{Liu et~al.(2022b)Liu, Perera, Khong, Chng, Styles and
  Khudanpur}]{liu2022efficient}
\bibinfo{author}{Liu, H.}, \bibinfo{author}{Perera, L.P.G.},
  \bibinfo{author}{Khong, A.W.}, \bibinfo{author}{Chng, E.S.},
  \bibinfo{author}{Styles, S.J.}, \bibinfo{author}{Khudanpur, S.},
  \bibinfo{year}{2022}b.
\newblock \bibinfo{title}{Efficient self-supervised learning representations
  for spoken language identification}.
\newblock \bibinfo{journal}{IEEE {J}ournal of {S}elected {T}opics in {S}ignal
  {P}rocessing} \bibinfo{volume}{16}, \bibinfo{pages}{1296--1307}.
%Type = Inproceedings
\bibitem[{Liu et~al.(2022c)Liu, Perera, Khong, Dauwels, Styles and
  Khudanpur}]{liu2022enhancing}
\bibinfo{author}{Liu, H.}, \bibinfo{author}{Perera, L.P.G.},
  \bibinfo{author}{Khong, A.W.}, \bibinfo{author}{Dauwels, J.},
  \bibinfo{author}{Styles, S.J.}, \bibinfo{author}{Khudanpur, S.},
  \bibinfo{year}{2022}c.
\newblock \bibinfo{title}{Enhancing language identification using dual-mode
  model with knowledge distillation}, in: \bibinfo{booktitle}{Odyssey: The
  Speaker and Language Recognition Workshop}, \bibinfo{organization}{ISCA}. pp.
  \bibinfo{pages}{248--254}.
%Type = Inproceedings
\bibitem[{Long et~al.(2015)Long, Cao, Wang and Jordan}]{long2015learning}
\bibinfo{author}{Long, M.}, \bibinfo{author}{Cao, Y.}, \bibinfo{author}{Wang,
  J.}, \bibinfo{author}{Jordan, M.}, \bibinfo{year}{2015}.
\newblock \bibinfo{title}{Learning transferable features with deep adaptation
  networks}, in: \bibinfo{booktitle}{International conference on machine
  learning}, \bibinfo{organization}{PMLR}. pp. \bibinfo{pages}{97--105}.
%Type = Article
\bibitem[{Lopez-Moreno et~al.(2016)Lopez-Moreno, Gonzalez-Dominguez, Martinez,
  Plchot, Gonzalez-Rodriguez and Moreno}]{lopez2016use}
\bibinfo{author}{Lopez-Moreno, I.}, \bibinfo{author}{Gonzalez-Dominguez, J.},
  \bibinfo{author}{Martinez, D.}, \bibinfo{author}{Plchot, O.},
  \bibinfo{author}{Gonzalez-Rodriguez, J.}, \bibinfo{author}{Moreno, P.J.},
  \bibinfo{year}{2016}.
\newblock \bibinfo{title}{On the use of deep feedforward neural networks for
  automatic language identification}.
\newblock \bibinfo{journal}{Computer {S}peech \& {L}anguage}
  \bibinfo{volume}{40}, \bibinfo{pages}{46--59}.
%Type = Inproceedings
\bibitem[{Loshchilov and Hutter(2018)}]{loshchilov2018decoupled}
\bibinfo{author}{Loshchilov, I.}, \bibinfo{author}{Hutter, F.},
  \bibinfo{year}{2018}.
\newblock \bibinfo{title}{Decoupled weight decay regularization}, in:
  \bibinfo{booktitle}{ICLR}.
%Type = Inproceedings
\bibitem[{Maity et~al.(2012)Maity, A.K., Rao and Nandi}]{maity2012iitkgp}
\bibinfo{author}{Maity, S.}, \bibinfo{author}{A.K., V.}, \bibinfo{author}{Rao,
  K.}, \bibinfo{author}{Nandi, D.}, \bibinfo{year}{2012}.
\newblock \bibinfo{title}{{IITKGP-MLILSC} speech database for language
  identification}, in: \bibinfo{booktitle}{{N}ational {C}onference on
  {C}ommunication (NCC)}, \bibinfo{organization}{IEEE}. pp.
  \bibinfo{pages}{1--5}.
%Type = Inproceedings
\bibitem[{Mandava et~al.(2019)Mandava, Vuddagiri, Vydana and
  Vuppala}]{mandava2019investigation}
\bibinfo{author}{Mandava, T.}, \bibinfo{author}{Vuddagiri, R.K.},
  \bibinfo{author}{Vydana, H.K.}, \bibinfo{author}{Vuppala, A.K.},
  \bibinfo{year}{2019}.
\newblock \bibinfo{title}{An investigation of {LSTM-CTC} based joint acoustic
  model for {I}ndian language identification}, in:
  \bibinfo{booktitle}{Automatic Speech Recognition and Understanding Workshop
  (ASRU)}, \bibinfo{organization}{IEEE}. pp. \bibinfo{pages}{389--396}.
%Type = Inproceedings
\bibitem[{Mandava and Vuppala(2019)}]{mandava2019attention}
\bibinfo{author}{Mandava, T.}, \bibinfo{author}{Vuppala, A.K.},
  \bibinfo{year}{2019}.
\newblock \bibinfo{title}{Attention based residual-time delay neural network
  for {I}ndian language identification}, in: \bibinfo{booktitle}{International
  Conference on Contemporary Computing (IC3)}, \bibinfo{organization}{IEEE}.
  pp. \bibinfo{pages}{1--5}.
%Type = Inproceedings
\bibitem[{Martinez et~al.(2011)Martinez, Plchot, Burget, Glembek and
  Mat{\v{e}}jka}]{martinez2011language}
\bibinfo{author}{Martinez, D.}, \bibinfo{author}{Plchot, O.},
  \bibinfo{author}{Burget, L.}, \bibinfo{author}{Glembek, O.},
  \bibinfo{author}{Mat{\v{e}}jka, P.}, \bibinfo{year}{2011}.
\newblock \bibinfo{title}{Language recognition in ivectors space}, in:
  \bibinfo{booktitle}{INTERSPEECH}, \bibinfo{organization}{ISCA}.
%Type = Inproceedings
\bibitem[{Mauch and Ewert(2013)}]{matthias2013a}
\bibinfo{author}{Mauch, M.}, \bibinfo{author}{Ewert, S.}, \bibinfo{year}{2013}.
\newblock \bibinfo{title}{The audio degradation toolbox and its application to
  robustness evaluation}, in: \bibinfo{booktitle}{International Society for
  Music Information Retrieval Conference (ISMIR)}, \bibinfo{address}{Curitiba,
  Brazil}.
%Type = Article
\bibitem[{Mohamed et~al.(2022)Mohamed, Lee, Borgholt, Havtorn, Edin, Igel,
  Kirchhoff, Li, Livescu, Maaløe, Sainath and Watanabe}]{mohamed2022self}
\bibinfo{author}{Mohamed, A.}, \bibinfo{author}{Lee, H.y.},
  \bibinfo{author}{Borgholt, L.}, \bibinfo{author}{Havtorn, J.D.},
  \bibinfo{author}{Edin, J.}, \bibinfo{author}{Igel, C.},
  \bibinfo{author}{Kirchhoff, K.}, \bibinfo{author}{Li, S.W.},
  \bibinfo{author}{Livescu, K.}, \bibinfo{author}{Maaløe, L.},
  \bibinfo{author}{Sainath, T.N.}, \bibinfo{author}{Watanabe, S.},
  \bibinfo{year}{2022}.
\newblock \bibinfo{title}{Self-supervised speech representation learning: A
  review}.
\newblock \bibinfo{journal}{{IEEE} Journal of {S}elected {T}opics in {S}ignal
  {P}rocessing} \bibinfo{volume}{16}, \bibinfo{pages}{1179--1210}.
%Type = Article
\bibitem[{Monteiro et~al.(2019)Monteiro, Alam and Falk}]{monteiro2019residual}
\bibinfo{author}{Monteiro, J.}, \bibinfo{author}{Alam, J.},
  \bibinfo{author}{Falk, T.H.}, \bibinfo{year}{2019}.
\newblock \bibinfo{title}{Residual convolutional neural network with attentive
  feature pooling for end-to-end language identification from short-duration
  speech}.
\newblock \bibinfo{journal}{Computer {S}peech \& {L}anguage}
  \bibinfo{volume}{58}, \bibinfo{pages}{364--376}.
%Type = Article
\bibitem[{Moreno-Torres et~al.(2012)Moreno-Torres, Raeder,
  Alaiz-Rodr{\'\i}guez, Chawla and Herrera}]{moreno2012unifying}
\bibinfo{author}{Moreno-Torres, J.G.}, \bibinfo{author}{Raeder, T.},
  \bibinfo{author}{Alaiz-Rodr{\'\i}guez, R.}, \bibinfo{author}{Chawla, N.V.},
  \bibinfo{author}{Herrera, F.}, \bibinfo{year}{2012}.
\newblock \bibinfo{title}{A unifying view on dataset shift in classification}.
\newblock \bibinfo{journal}{Pattern {R}ecognition} \bibinfo{volume}{45},
  \bibinfo{pages}{521--530}.
%Type = Misc
\bibitem[{Mozilla()}]{mozilla}
\bibinfo{author}{Mozilla}, .
\newblock \bibinfo{title}{{Common Voice}}.
\newblock \bibinfo{howpublished}{\url{https://commonvoice.mozilla.org/en}}.
\newblock \bibinfo{note}{Accessed: 2020-01-12}.
%Type = Article
\bibitem[{Mushtaq and Su(2020)}]{mushtaq2020environmental}
\bibinfo{author}{Mushtaq, Z.}, \bibinfo{author}{Su, S.F.},
  \bibinfo{year}{2020}.
\newblock \bibinfo{title}{Environmental sound classification using a
  regularized deep convolutional neural network with data augmentation}.
\newblock \bibinfo{journal}{Applied {A}coustics} \bibinfo{volume}{167},
  \bibinfo{pages}{107389}.
%Type = Inproceedings
\bibitem[{Nadimpalli and Rattani(2022)}]{nadimpalli2022improving}
\bibinfo{author}{Nadimpalli, A.V.}, \bibinfo{author}{Rattani, A.},
  \bibinfo{year}{2022}.
\newblock \bibinfo{title}{On improving cross-dataset generalization of deepfake
  detectors}, in: \bibinfo{booktitle}{CVPR}, \bibinfo{organization}{IEEE/CVF}.
  pp. \bibinfo{pages}{91--99}.
%Type = Article
\bibitem[{Padi et~al.(2020)Padi, Mohan and Ganapathy}]{padi2020towards}
\bibinfo{author}{Padi, B.}, \bibinfo{author}{Mohan, A.},
  \bibinfo{author}{Ganapathy, S.}, \bibinfo{year}{2020}.
\newblock \bibinfo{title}{Towards relevance and sequence modeling in language
  recognition}.
\newblock \bibinfo{journal}{{IEEE/ACM} Transactions on Audio, Speech, and
  Language Processing} \bibinfo{volume}{28}, \bibinfo{pages}{1223--1232}.
%Type = Article
\bibitem[{Pan et~al.(2010)Pan, Tsang, Kwok and Yang}]{pan2010domain}
\bibinfo{author}{Pan, S.J.}, \bibinfo{author}{Tsang, I.W.},
  \bibinfo{author}{Kwok, J.T.}, \bibinfo{author}{Yang, Q.},
  \bibinfo{year}{2010}.
\newblock \bibinfo{title}{Domain adaptation via transfer component analysis}.
\newblock \bibinfo{journal}{IEEE {T}ransactions on {N}eural {N}etworks}
  \bibinfo{volume}{22}, \bibinfo{pages}{199--210}.
%Type = Article
\bibitem[{Pandey and Wang(2022)}]{pandeyself2022}
\bibinfo{author}{Pandey, A.}, \bibinfo{author}{Wang, D.}, \bibinfo{year}{2022}.
\newblock \bibinfo{title}{Self-attending {RNN} for speech enhancement to
  improve cross-corpus generalization}.
\newblock \bibinfo{journal}{IEEE/ACM {T}ransactions on {A}udio, {S}peech, and
  {L}anguage {P}rocessing} \bibinfo{volume}{30}, \bibinfo{pages}{1374--1385}.
%Type = Inproceedings
\bibitem[{Park et~al.(2019)}]{park2019specaugment}
\bibinfo{author}{Park, D.S.}, et~al., \bibinfo{year}{2019}.
\newblock \bibinfo{title}{Spec{A}ugment: A simple data augmentation method for
  automatic speech recognition}, in: \bibinfo{booktitle}{INTERSPEECH},
  \bibinfo{organization}{ISCA}. pp. \bibinfo{pages}{2613--2617}.
%Type = Inproceedings
\bibitem[{Paszke et~al.(2019)Paszke, Gross, Massa and
  Lerer}]{paszke2019pytorch}
\bibinfo{author}{Paszke, A.}, \bibinfo{author}{Gross, S.},
  \bibinfo{author}{Massa, F.}, \bibinfo{author}{Lerer, A.},
  \bibinfo{year}{2019}.
\newblock \bibinfo{title}{{P}y{T}orch: An imperative style, high-performance
  deep learning library.}, in: \bibinfo{booktitle}{NeurIPS}, pp.
  \bibinfo{pages}{8024--8035}.
%Type = Inproceedings
\bibitem[{Paul et~al.(2017)Paul, Sahidullah and Saha}]{paul2017}
\bibinfo{author}{Paul, D.}, \bibinfo{author}{Sahidullah, M.},
  \bibinfo{author}{Saha, G.}, \bibinfo{year}{2017}.
\newblock \bibinfo{title}{Generalization of spoofing countermeasures: A case
  study with {ASV}spoof 2015 and {BTAS} 2016 corpora}, in:
  \bibinfo{booktitle}{ICASSP}, \bibinfo{organization}{IEEE}. pp.
  \bibinfo{pages}{2047--2051}.
%Type = Inproceedings
\bibitem[{Povey et~al.(2011)}]{povey2011kaldi}
\bibinfo{author}{Povey, D.}, et~al., \bibinfo{year}{2011}.
\newblock \bibinfo{title}{The {K}aldi speech recognition toolkit}, in:
  \bibinfo{booktitle}{Automatic Speech Recognition and Understanding Workshop
  (ASRU)}, \bibinfo{organization}{IEEE}.
%Type = Article
\bibitem[{Radford et~al.(2022)Radford, Kim, Xu, Brockman, McLeavey and
  Sutskever}]{radford2022robust}
\bibinfo{author}{Radford, A.}, \bibinfo{author}{Kim, J.W.},
  \bibinfo{author}{Xu, T.}, \bibinfo{author}{Brockman, G.},
  \bibinfo{author}{McLeavey, C.}, \bibinfo{author}{Sutskever, I.},
  \bibinfo{year}{2022}.
\newblock \bibinfo{title}{Robust speech recognition via large-scale weak
  supervision}.
\newblock \bibinfo{journal}{OpenAI Blog} .
%Type = Article
\bibitem[{Reddy et~al.(2013)Reddy, Maity and Rao}]{reddy2013identification}
\bibinfo{author}{Reddy, V.R.}, \bibinfo{author}{Maity, S.},
  \bibinfo{author}{Rao, K.S.}, \bibinfo{year}{2013}.
\newblock \bibinfo{title}{Identification of {I}ndian languages using
  multi-level spectral and prosodic features}.
\newblock \bibinfo{journal}{International Journal of Speech Technology}
  \bibinfo{volume}{16}, \bibinfo{pages}{489--511}.
%Type = Inproceedings
\bibitem[{Ribas et~al.(2016)Ribas, Vincent and Calvo}]{ribas2016study}
\bibinfo{author}{Ribas, D.}, \bibinfo{author}{Vincent, E.},
  \bibinfo{author}{Calvo, J.R.}, \bibinfo{year}{2016}.
\newblock \bibinfo{title}{A study of speech distortion conditions in real
  scenarios for speech processing applications}, in: \bibinfo{booktitle}{Spoken
  Language Technology Workshop (SLT)}, pp. \bibinfo{pages}{13--20}.
%Type = Inproceedings
\bibitem[{Rossenbach et~al.(2020)Rossenbach, Zeyer, Schl{\"u}ter and
  Ney}]{rossenbach2020generating}
\bibinfo{author}{Rossenbach, N.}, \bibinfo{author}{Zeyer, A.},
  \bibinfo{author}{Schl{\"u}ter, R.}, \bibinfo{author}{Ney, H.},
  \bibinfo{year}{2020}.
\newblock \bibinfo{title}{Generating synthetic audio data for attention-based
  speech recognition systems}, in: \bibinfo{booktitle}{ICASSP},
  \bibinfo{organization}{IEEE}. pp. \bibinfo{pages}{7069--7073}.
%Type = Article
\bibitem[{Ruder(2017)}]{ruder2017overview}
\bibinfo{author}{Ruder, S.}, \bibinfo{year}{2017}.
\newblock \bibinfo{title}{An overview of multi-task learning in deep neural
  networks}.
\newblock \bibinfo{journal}{arXiv preprint arXiv:1706.05098} .
%Type = Inproceedings
\bibitem[{Sadjadi et~al.(2018)Sadjadi, Kheyrkhah, Tong, Greenberg, Reynolds,
  Singer, Mason and Hernandez-Cordero}]{sadjadi20182017}
\bibinfo{author}{Sadjadi, S.O.}, \bibinfo{author}{Kheyrkhah, T.},
  \bibinfo{author}{Tong, A.}, \bibinfo{author}{Greenberg, C.S.},
  \bibinfo{author}{Reynolds, D.A.}, \bibinfo{author}{Singer, E.},
  \bibinfo{author}{Mason, L.P.}, \bibinfo{author}{Hernandez-Cordero, J.},
  \bibinfo{year}{2018}.
\newblock \bibinfo{title}{The 2017 {NIST} language recognition evaluation.},
  in: \bibinfo{booktitle}{Odyssey: The Speaker and Language Recognition
  Workshop}, \bibinfo{organization}{ISCA}.
%Type = Article
\bibitem[{Salamon and Bello(2017)}]{salamon2017deep}
\bibinfo{author}{Salamon, J.}, \bibinfo{author}{Bello, J.P.},
  \bibinfo{year}{2017}.
\newblock \bibinfo{title}{Deep convolutional neural networks and data
  augmentation for environmental sound classification}.
\newblock \bibinfo{journal}{IEEE {S}ignal {P}rocessing {L}etters}
  \bibinfo{volume}{24}, \bibinfo{pages}{279--283}.
%Type = Inproceedings
\bibitem[{Sarfjoo et~al.(2020)Sarfjoo, Madikeri, Motlicek and
  Marcel}]{sarfjoo20_interspeech}
\bibinfo{author}{Sarfjoo, S.}, \bibinfo{author}{Madikeri, S.},
  \bibinfo{author}{Motlicek, P.}, \bibinfo{author}{Marcel, S.},
  \bibinfo{year}{2020}.
\newblock \bibinfo{title}{Supervised domain adaptation for text-independent
  speaker verification using limited data}, in:
  \bibinfo{booktitle}{INTERSPEECH}, pp. \bibinfo{pages}{3815--3819}.
%Type = Article
\bibitem[{Schuller et~al.(2010)Schuller, Vlasenko, Eyben, W{\"o}llmer,
  Stuhlsatz, Wendemuth and Rigoll}]{schuller2010cross}
\bibinfo{author}{Schuller, B.}, \bibinfo{author}{Vlasenko, B.},
  \bibinfo{author}{Eyben, F.}, \bibinfo{author}{W{\"o}llmer, M.},
  \bibinfo{author}{Stuhlsatz, A.}, \bibinfo{author}{Wendemuth, A.},
  \bibinfo{author}{Rigoll, G.}, \bibinfo{year}{2010}.
\newblock \bibinfo{title}{Cross-corpus acoustic emotion recognition: Variances
  and strategies}.
\newblock \bibinfo{journal}{IEEE {T}ransactions on {A}ffective {C}omputing}
  \bibinfo{volume}{1}, \bibinfo{pages}{119--131}.
%Type = Inproceedings
\bibitem[{Shen et~al.(2017)Shen, Lu, Li and Kawai}]{shen2017conditional}
\bibinfo{author}{Shen, P.}, \bibinfo{author}{Lu, X.}, \bibinfo{author}{Li, S.},
  \bibinfo{author}{Kawai, H.}, \bibinfo{year}{2017}.
\newblock \bibinfo{title}{Conditional generative adversarial nets classifier
  for spoken language identification.}, in: \bibinfo{booktitle}{INTERSPEECH},
  \bibinfo{organization}{ISCA}. pp. \bibinfo{pages}{2814--2818}.
%Type = Inproceedings
\bibitem[{Singh et~al.(2021)Singh, Saha and Sahidullah}]{singh2021non}
\bibinfo{author}{Singh, P.}, \bibinfo{author}{Saha, G.},
  \bibinfo{author}{Sahidullah, M.}, \bibinfo{year}{2021}.
\newblock \bibinfo{title}{Non-linear frequency warping using constant-{Q}
  transformation for speech emotion recognition}, in:
  \bibinfo{booktitle}{International Conference on Computer Communication and
  Informatics (ICCCI)}, \bibinfo{organization}{IEEE}. pp.
  \bibinfo{pages}{1--6}.
%Type = Article
\bibitem[{Snyder et~al.(2015)Snyder, Chen and Povey}]{snyder2015musan}
\bibinfo{author}{Snyder, D.}, \bibinfo{author}{Chen, G.},
  \bibinfo{author}{Povey, D.}, \bibinfo{year}{2015}.
\newblock \bibinfo{title}{{MUSAN}: A music, speech, and noise corpus}.
\newblock \bibinfo{journal}{arXiv preprint arXiv:1510.08484} .
%Type = Inproceedings
\bibitem[{Snyder et~al.(2018a)}]{snyder2018spoken}
\bibinfo{author}{Snyder, D.}, et~al., \bibinfo{year}{2018}a.
\newblock \bibinfo{title}{Spoken language recognition using x-vectors.}, in:
  \bibinfo{booktitle}{Odyssey: The Speaker and Language Recognition Workshop},
  pp. \bibinfo{pages}{105--111}.
%Type = Inproceedings
\bibitem[{Snyder et~al.(2018b)}]{snyder2018x}
\bibinfo{author}{Snyder, D.}, et~al., \bibinfo{year}{2018}b.
\newblock \bibinfo{title}{X-vectors: Robust {DNN} embeddings for speaker
  recognition}, in: \bibinfo{booktitle}{ICASSP}, \bibinfo{organization}{IEEE}.
  pp. \bibinfo{pages}{5329--5333}.
%Type = Article
\bibitem[{Sturm(2014)}]{sturm2014simple}
\bibinfo{author}{Sturm, B.L.}, \bibinfo{year}{2014}.
\newblock \bibinfo{title}{A simple method to determine if a music information
  retrieval system is a “horse”}.
\newblock \bibinfo{journal}{IEEE {T}ransactions on {M}ultimedia}
  \bibinfo{volume}{16}, \bibinfo{pages}{1636--1644}.
%Type = Inproceedings
\bibitem[{Tang et~al.(2019)Tang, Wang and Song}]{tang2019ap19}
\bibinfo{author}{Tang, Z.}, \bibinfo{author}{Wang, D.}, \bibinfo{author}{Song,
  L.}, \bibinfo{year}{2019}.
\newblock \bibinfo{title}{{AP19}-{OLR} challenge: Three tasks and their
  baselines}, in: \bibinfo{booktitle}{Asia-{P}acific Signal and Information
  Processing Association Annual Summit and Conference (APSIPA ASC)},
  \bibinfo{organization}{IEEE}. pp. \bibinfo{pages}{1917--1921}.
%Type = Inproceedings
\bibitem[{Thienpondt et~al.(2022)Thienpondt, Desplanques and
  Demuynck}]{thienpondt2022tackling}
\bibinfo{author}{Thienpondt, J.}, \bibinfo{author}{Desplanques, B.},
  \bibinfo{author}{Demuynck, K.}, \bibinfo{year}{2022}.
\newblock \bibinfo{title}{Tackling the score shift in cross-lingual speaker
  verification by exploiting language information}, in:
  \bibinfo{booktitle}{ICASSP}, \bibinfo{organization}{IEEE}. pp.
  \bibinfo{pages}{7187--7191}.
%Type = Inproceedings
\bibitem[{Toledo-Ronen and Sorin(2013)}]{toledo2013voice}
\bibinfo{author}{Toledo-Ronen, O.}, \bibinfo{author}{Sorin, A.},
  \bibinfo{year}{2013}.
\newblock \bibinfo{title}{Voice-based sadness and anger recognition with
  cross-corpora evaluation}, in: \bibinfo{booktitle}{ICASSP},
  \bibinfo{organization}{IEEE}. pp. \bibinfo{pages}{7517--7521}.
%Type = Inproceedings
\bibitem[{Tong et~al.(2021)Tong, Zhao, Zhou, Lu, Li, Li and Hong}]{tong2021asv}
\bibinfo{author}{Tong, F.}, \bibinfo{author}{Zhao, M.}, \bibinfo{author}{Zhou,
  J.}, \bibinfo{author}{Lu, H.}, \bibinfo{author}{Li, Z.}, \bibinfo{author}{Li,
  L.}, \bibinfo{author}{Hong, Q.}, \bibinfo{year}{2021}.
\newblock \bibinfo{title}{{ASV}-{S}ubtools: Open source toolkit for automatic
  speaker verification}, in: \bibinfo{booktitle}{ICASSP},
  \bibinfo{organization}{IEEE}. pp. \bibinfo{pages}{6184--6188}.
%Type = Inproceedings
\bibitem[{Tsakalidis and Byrne(2005)}]{tsakalidis2005acoustic}
\bibinfo{author}{Tsakalidis, S.}, \bibinfo{author}{Byrne, W.},
  \bibinfo{year}{2005}.
\newblock \bibinfo{title}{Acoustic training from heterogeneous data sources:
  Experiments in {M}andarin conversational telephone speech transcription}, in:
  \bibinfo{booktitle}{ICASSP}, \bibinfo{organization}{IEEE}. pp.
  \bibinfo{pages}{I--461}.
%Type = Inproceedings
\bibitem[{Valk and Alum{\"a}e(2021)}]{valk2021voxlingua107}
\bibinfo{author}{Valk, J.}, \bibinfo{author}{Alum{\"a}e, T.},
  \bibinfo{year}{2021}.
\newblock \bibinfo{title}{Vox{L}ingua107: a dataset for spoken language
  recognition}, in: \bibinfo{booktitle}{Spoken Language Technology Workshop
  (SLT)}, \bibinfo{organization}{IEEE}. pp. \bibinfo{pages}{652--658}.
%Type = Inproceedings
\bibitem[{Vlasenko et~al.(2013)Vlasenko, Philippou-H{\"u}bner and
  Wendemuth}]{vlasenko2013parameter}
\bibinfo{author}{Vlasenko, B.}, \bibinfo{author}{Philippou-H{\"u}bner, D.},
  \bibinfo{author}{Wendemuth, A.}, \bibinfo{year}{2013}.
\newblock \bibinfo{title}{Parameter optimization issues for cross-corpora
  emotion classification}, in: \bibinfo{booktitle}{Humaine Association
  Conference on Affective Computing and Intelligent Interaction},
  \bibinfo{organization}{IEEE}. pp. \bibinfo{pages}{454--459}.
%Type = Article
\bibitem[{Vlasenko et~al.(2014)Vlasenko, Prylipko, B{\"o}ck and
  Wendemuth}]{vlasenko2014modeling}
\bibinfo{author}{Vlasenko, B.}, \bibinfo{author}{Prylipko, D.},
  \bibinfo{author}{B{\"o}ck, R.}, \bibinfo{author}{Wendemuth, A.},
  \bibinfo{year}{2014}.
\newblock \bibinfo{title}{Modeling phonetic pattern variability in favor of the
  creation of robust emotion classifiers for real-life applications}.
\newblock \bibinfo{journal}{Computer {S}peech \& {L}anguage}
  \bibinfo{volume}{28}, \bibinfo{pages}{483--500}.
%Type = Inproceedings
\bibitem[{Vuddagiri et~al.(2018)}]{vuddagiri2018iiith}
\bibinfo{author}{Vuddagiri}, et~al., \bibinfo{year}{2018}.
\newblock \bibinfo{title}{{IIITH-ILSC} speech database for {I}ndain language
  identification.}, in: \bibinfo{booktitle}{Spoken Language Technologies for
  Under-Resourced Languages}, \bibinfo{organization}{IEEE}. pp.
  \bibinfo{pages}{56--60}.
%Type = Article
\bibitem[{Wang et~al.(2018a)Wang, Cheng, Liu and Liu}]{wang2018additive}
\bibinfo{author}{Wang, F.}, \bibinfo{author}{Cheng, J.}, \bibinfo{author}{Liu,
  W.}, \bibinfo{author}{Liu, H.}, \bibinfo{year}{2018}a.
\newblock \bibinfo{title}{Additive margin softmax for face verification}.
\newblock \bibinfo{journal}{IEEE {S}ignal {P}rocessing {L}etters}
  \bibinfo{volume}{25}, \bibinfo{pages}{926--930}.
%Type = Inproceedings
\bibitem[{Wang et~al.(2018b)Wang, Zhu, Gong and Li}]{wang2018transferable}
\bibinfo{author}{Wang, J.}, \bibinfo{author}{Zhu, X.}, \bibinfo{author}{Gong,
  S.}, \bibinfo{author}{Li, W.}, \bibinfo{year}{2018}b.
\newblock \bibinfo{title}{Transferable joint attribute-identity deep learning
  for unsupervised person re-identification}, in: \bibinfo{booktitle}{CVPR},
  \bibinfo{organization}{IEEE}. pp. \bibinfo{pages}{2275--2284}.
%Type = Article
\bibitem[{Wang and Deng(2018)}]{wang2018deep}
\bibinfo{author}{Wang, M.}, \bibinfo{author}{Deng, W.}, \bibinfo{year}{2018}.
\newblock \bibinfo{title}{Deep visual domain adaptation: A survey}.
\newblock \bibinfo{journal}{Neurocomputing} \bibinfo{volume}{312},
  \bibinfo{pages}{135--153}.
%Type = Inproceedings
\bibitem[{Wei et~al.(2020)Wei, Zou, Liao et~al.}]{wei2020comparison}
\bibinfo{author}{Wei, S.}, \bibinfo{author}{Zou, S.}, \bibinfo{author}{Liao,
  F.}, et~al., \bibinfo{year}{2020}.
\newblock \bibinfo{title}{A comparison on data augmentation methods based on
  deep learning for audio classification}, in: \bibinfo{booktitle}{Journal of
  {P}hysics: Conference Series}, \bibinfo{organization}{IOP Publishing}.
%Type = Inproceedings
\bibitem[{Xia et~al.(2021)Xia, Zhang, Weng, Yu and Yu}]{xia2021self}
\bibinfo{author}{Xia, W.}, \bibinfo{author}{Zhang, C.}, \bibinfo{author}{Weng,
  C.}, \bibinfo{author}{Yu, M.}, \bibinfo{author}{Yu, D.},
  \bibinfo{year}{2021}.
\newblock \bibinfo{title}{Self-supervised text-independent speaker verification
  using prototypical momentum contrastive learning}, in:
  \bibinfo{booktitle}{ICASSP}, \bibinfo{organization}{IEEE}. pp.
  \bibinfo{pages}{6723--6727}.
%Type = Inproceedings
\bibitem[{Yan et~al.(2017)Yan, Ding, Li, Wang, Xu and Zuo}]{yan2017mind}
\bibinfo{author}{Yan, H.}, \bibinfo{author}{Ding, Y.}, \bibinfo{author}{Li,
  P.}, \bibinfo{author}{Wang, Q.}, \bibinfo{author}{Xu, Y.},
  \bibinfo{author}{Zuo, W.}, \bibinfo{year}{2017}.
\newblock \bibinfo{title}{Mind the class weight bias: Weighted maximum mean
  discrepancy for unsupervised domain adaptation}, in:
  \bibinfo{booktitle}{CVPR}, \bibinfo{organization}{IEEE}. pp.
  \bibinfo{pages}{2272--2281}.
%Type = Inproceedings
\bibitem[{Zhang et~al.(2018)Zhang, Cisse, Dauphin and
  Lopez-Paz}]{zhang2017mixup}
\bibinfo{author}{Zhang, H.}, \bibinfo{author}{Cisse, M.},
  \bibinfo{author}{Dauphin, Y.N.}, \bibinfo{author}{Lopez-Paz, D.},
  \bibinfo{year}{2018}.
\newblock \bibinfo{title}{mixup: Beyond empirical risk minimization}, in:
  \bibinfo{booktitle}{ICLR}.
%Type = Article
\bibitem[{Zhang et~al.(2020)Zhang, Deng, Tang, Zhang and
  Jia}]{zhang2020unsupervised}
\bibinfo{author}{Zhang, Y.}, \bibinfo{author}{Deng, B.}, \bibinfo{author}{Tang,
  H.}, \bibinfo{author}{Zhang, L.}, \bibinfo{author}{Jia, K.},
  \bibinfo{year}{2020}.
\newblock \bibinfo{title}{Unsupervised multi-class domain adaptation: Theory,
  algorithms, and practice}.
\newblock \bibinfo{journal}{IEEE {T}ransactions on {P}attern {A}nalysis and
  {M}achine {I}ntelligence} .
%Type = Article
\bibitem[{Zhang and Yang(2021)}]{zhang2021survey}
\bibinfo{author}{Zhang, Y.}, \bibinfo{author}{Yang, Q.}, \bibinfo{year}{2021}.
\newblock \bibinfo{title}{A survey on multi-task learning}.
\newblock \bibinfo{journal}{IEEE {T}ransactions on {K}nowledge and {D}ata
  {E}ngineering} , \bibinfo{pages}{1--1}.
%Type = Inproceedings
\bibitem[{Zhang et~al.(2011)Zhang, Weninger, W{\"o}llmer and
  Schuller}]{zhang2011unsupervised}
\bibinfo{author}{Zhang, Z.}, \bibinfo{author}{Weninger, F.},
  \bibinfo{author}{W{\"o}llmer, M.}, \bibinfo{author}{Schuller, B.},
  \bibinfo{year}{2011}.
\newblock \bibinfo{title}{Unsupervised learning in cross-corpus acoustic
  emotion recognition}, in: \bibinfo{booktitle}{Automatic Speech Recognition
  and Understanding Workshop (ASRU)}, \bibinfo{organization}{IEEE}. pp.
  \bibinfo{pages}{523--528}.
%Type = Article
\bibitem[{Zhou et~al.(2022)Zhou, Liu, Qiao, Xiang and Loy}]{zhou2022domain}
\bibinfo{author}{Zhou, K.}, \bibinfo{author}{Liu, Z.}, \bibinfo{author}{Qiao,
  Y.}, \bibinfo{author}{Xiang, T.}, \bibinfo{author}{Loy, C.C.},
  \bibinfo{year}{2022}.
\newblock \bibinfo{title}{Domain generalization: {A} survey}.
\newblock \bibinfo{journal}{IEEE {T}ransactions on {P}attern {A}nalysis and
  {M}achine {I}ntelligence} .
%Type = Inproceedings
\bibitem[{Zhu et~al.(2015)Zhu, Zhang, Zhao and Wu}]{zhu2015transfer}
\bibinfo{author}{Zhu, R.}, \bibinfo{author}{Zhang, T.}, \bibinfo{author}{Zhao,
  Q.}, \bibinfo{author}{Wu, Z.}, \bibinfo{year}{2015}.
\newblock \bibinfo{title}{A transfer learning approach to cross-database facial
  expression recognition}, in: \bibinfo{booktitle}{International Conference on
  Biometrics (ICB)}, \bibinfo{organization}{IEEE}. pp.
  \bibinfo{pages}{293--298}.
%Type = Article
\bibitem[{Zhu et~al.(2020)Zhu, Zhuang, Wang, Ke, Chen, Bian, Xiong and
  He}]{zhu2020deep}
\bibinfo{author}{Zhu, Y.}, \bibinfo{author}{Zhuang, F.}, \bibinfo{author}{Wang,
  J.}, \bibinfo{author}{Ke, G.}, \bibinfo{author}{Chen, J.},
  \bibinfo{author}{Bian, J.}, \bibinfo{author}{Xiong, H.}, \bibinfo{author}{He,
  Q.}, \bibinfo{year}{2020}.
\newblock \bibinfo{title}{Deep subdomain adaptation network for image
  classification}.
\newblock \bibinfo{journal}{IEEE {T}ransactions on {N}eural {N}etworks and
  {L}earning {S}ystems} \bibinfo{volume}{32}, \bibinfo{pages}{1713--1722}.
%Type = Article
\bibitem[{Zhuang et~al.(2020)Zhuang, Qi, Duan, Xi, Zhu, Zhu, Xiong and
  He}]{zhuang2020comprehensive}
\bibinfo{author}{Zhuang, F.}, \bibinfo{author}{Qi, Z.}, \bibinfo{author}{Duan,
  K.}, \bibinfo{author}{Xi, D.}, \bibinfo{author}{Zhu, Y.},
  \bibinfo{author}{Zhu, H.}, \bibinfo{author}{Xiong, H.}, \bibinfo{author}{He,
  Q.}, \bibinfo{year}{2020}.
\newblock \bibinfo{title}{A comprehensive survey on transfer learning}.
\newblock \bibinfo{journal}{Proceedings of the IEEE} \bibinfo{volume}{109},
  \bibinfo{pages}{43--76}.

\end{thebibliography}

\end{document}